\documentclass[sigconf]{acmart}

\usepackage{amsfonts}
\usepackage{bm}
\usepackage{subcaption}
\usepackage{mathtools}
\usepackage{hyperref}

\AtBeginDocument{%
  \providecommand\BibTeX{{%
    \normalfont B\kern-0.5em{\scshape i\kern-0.25em b}\kern-0.8em\TeX}}}

\copyrightyear{2022} 
\acmYear{2022} 
\setcopyright{acmcopyright}\acmConference[SIGIR '22]{Proceedings of the 45th International ACM SIGIR Conference on Research and Development in Information Retrieval}{July 11--15, 2022}{Madrid, Spain}
\acmBooktitle{Proceedings of the 45th International ACM SIGIR Conference on Research and Development in Information Retrieval (SIGIR '22), July 11--15, 2022, Madrid, Spain}
\acmPrice{15.00}
\acmDOI{10.1145/3477495.3532072}
\acmISBN{978-1-4503-8732-3/22/07}

\begin{document}
\fancyhead{}
\title{Unsupervised Belief Representation Learning with Information-Theoretic Variational Graph Auto-Encoders}

\author{
    Jinning Li\textsuperscript{\rm 1},
    Huajie Shao\textsuperscript{\rm 2},
    Dachun Sun\textsuperscript{\rm 1},
    Ruijie Wang\textsuperscript{\rm 1},
    Yuchen Yan\textsuperscript{\rm 1},
    Jinyang Li\textsuperscript{\rm 1},
    Shengzhong Liu\textsuperscript{\rm 1},\\
    Hanghang Tong\textsuperscript{\rm 1}, and
    Tarek Abdelzaher\textsuperscript{\rm 1}
}
\affiliation{
    \institution{\textsuperscript{\rm 1} Department of Computer Science, University of Illinois Urbana-Champaign}
    \country{}
}
\affiliation{
    \institution{\textsuperscript{\rm 2} Department of Computer Science, College of William and Mary}
    \country{}
}
\email{jinning4@illinois.edu, hshao@wm.edu, {dsun18, ruijiew2, yucheny5, jinyang7, sl29, htong, zaher}@illinois.edu}

\renewcommand{\shortauthors}{Li, et al.}



\begin{abstract}
This paper develops a novel unsupervised algorithm for {\em belief representation learning\/} in polarized networks that (i) uncovers the latent dimensions of the underlying belief space and (ii) {\em jointly embeds\/} users and content items (that they interact with) into that space in a manner that facilitates a number of downstream tasks, such as stance detection, stance prediction, and ideology mapping. Inspired by total correlation in information theory, we propose the Information-Theoretic Variational Graph Auto-Encoder (InfoVGAE) that learns to project both users and content items (e.g., posts that represent user views) into an appropriate disentangled latent space. To better disentangle latent variables in that space, we develop a total correlation regularization module, a Proportional-Integral (PI) control module, and adopt rectified Gaussian distribution to ensure the orthogonality.
The latent representation of users and content can then be used to quantify their ideological leaning and detect/predict their stances on issues. We evaluate the performance of the proposed InfoVGAE on three real-world datasets, of which two are collected from Twitter and one from U.S. Congress voting records. 
The evaluation results show that our model outperforms state-of-the-art unsupervised models by reducing $10.5\%$ user clustering errors and achieving $12.1\%$ higher F1 scores for stance separation of content items.
In addition, InfoVGAE produces a comparable result with supervised models. We also discuss its performance on stance prediction and user ranking within ideological groups.
\end{abstract}




\maketitle
\vspace{-8pt}
\section{Introduction}
The goal of this work is to develop an {\em unsupervised technique\/} for learning the latent structures that reveal human stances, viewpoints, or ideological preferences. We call this latent space discovery problem {\em unsupervised belief representation learning\/}. Applications of this work include (i) unsupervised {\em stance detection\/} (i.e., interpreting the stance espoused by a user or content item on a topic), (ii) unsupervised {\em stance prediction\/} (i.e., predicting user stance on a topic in the absence of a direct observation), and (iii) unsupervised polarity identification (i.e., recognizing the ideology most aligned with a particular group of users or statements).  

In this paper, we refer by
{\em stance\/} to a position on {\em one topic\/}. A stance detection algorithm~\cite{aldayel2019your,kuccuk2020stance}, for instance, might determine that the sentence ``I believe the president should resign" is of {\em negative stance\/} on the president. In contrast, by {\em polarity\/}, we refer to an ideological leaning, which could separate two ideologies or systems of belief. For example, in the US, Democrats and Republicans might disagree on a range of issues such as gun control, abortion rights, immigration, defense funding, border control infrastructure, and other topics. Once an individual is identified as of, say, a Republican-leaning (i.e., a Republican {\em polarity\/}), this polarity automatically predicts their possible stances on a multitude of topics. Belief representation learning allows us to jointly identify (i) which stances go with which polarity (i.e., with which side of the divide), and (ii) the polarity of different individual users. 

Most existing approaches formulate ideology learning as entity classification or link prediction tasks~\cite{hasan2014you,darwish2017improved,jiang2021social,kuccuk2020stance,wang2019heterogeneous,xiao2020timme,CMH}, but largely neglect the interpretability of learned representation space.
In contrast, our method automatically maps users and content items into an explainable latent space by disentangling orthogonal axes to represent the diversity of latent ideologies. Notably, the discovery of the underlying latent axes occurs in a complete {\em unsupervised\/} fashion. 
Intuitively, if a society is broken down into groups where members of the same group have coinciding stances on multitudes of issues (e.g., it might be that groups who are pro-choice also typically favor stricter gun control and more environmental spending), a lower-dimensional representation of the latent belief space becomes possible. Both the users and content items (e.g., tweets) are mapped to points in that belief space. The coordinates of a point represent its relative adherence with the respective latent ideologies. We call the problem of identifying dimensions of the latent space {\em belief representation learning\/} and call the mapping of users and content items into that space {\em belief embedding\/}. Note that, beliefs espoused by different polarities are not necessarily ``opposite". Rather, they may represent alternatives, such as supporting military spending versus environmental spending. Some individuals might indeed combine elements of different ideologies. Thus, the latent representation also allows describing individuals with mixed beliefs (e.g., who approve both military and environmental spending).


Our solution first constructs a bipartite heterogeneous information graph of users and the content items (we henceforth call {\em claims\/}), which has been shown effective to jointly model node attributes and local topology~\cite{kipf2016semi,magdy2016isisisnotislam,sridhar2015joint,shao2020mis,wang2018ace,wang2022}. We propose an unsupervised Information-Theoretic Variational Graph Autoencoder (InfoVGAE) to encode both users and claims in the graph into the same disentangled latent space. The disentangled and explainable property benefits the social analysis of users and claims, and boosts many downstream tasks. We develop several techniques to ensure this property. (i) We use a Rectified Gaussian Distribution to constrain the distribution of latent space to non-negative values. As we well explain later, this adjustment produces more explainable latent representations that are better aligned with our notion of latent beliefs. (ii) Inspired by information theory, we develop a total correlation regularizer to minimize the total correlation across different dimensions of latent representations. This forces the model to learn a statistically independent axis system. (iii) We use a Proportional-Integral (PI) controller~\cite{shao2020controlvae} from control theory to manipulate the value of KL-divergence to balance the trade-off between reconstruction quality and disentanglement.

We evaluate the performance of the proposed InfoVGAE on two real-world Twitter datasets (one about the 2020 US presidential election and one about the Eurovision 2016 song contest) and a Congress bills dataset collected from the Voteview database. We publish the source code and datasets as an additional contribution\footnote{The code and datasets are available at \url{https://github.com/jinningli/InfoVGAE}}.
The evaluation results illustrate that our method outperforms the state-of-the-art unsupervised dimensional reduction methods, graph embedding models, and stance detection models on all three datasets and produces a comparable result with supervised models. 
On Twitter datasets, our model outperforms the best baseline by reducing $7.1\%$ user polarity clustering errors, and achieving $13.4\%$ higher F1 score for tweet stance separation. On the Voteview dataset, our model reduces Congress member clustering errors by $17.4\%$, and improves F1 scores by $10.8\%$ for bill ideology separation.
In addition, the gap of F1 measure between supervised models and unsupervised models is further narrowed into only $1.86\%$ and $0.06\%$ on Twitter and Voteview dataset. Ablation studies are also conducted to validate the effectiveness of proposed components. We also discuss stance prediction and show user ranking within ideological groups, demonstrating a match with ground truth.

\section{Unsupervised Belief Representation Learning}\label{sec:formulation}
In this section, we first define the problem of unsupervised belief representation learning. Then, we introduce the way to build the Bipartite Heterogeneous Information Network (BHIN) and present the proposed Variational Graph Auto-Encoder method (InfoVGAE).

\subsection{Problem Statement}
Consider a set, $S$, of {\em users\/}, who collectively act on a set of content items, $C$, we call {\em claims\/} (a claim generally espouses a statement of belief or a view on issues). Examples of such content items could be tweets on Twitter, posts on Reddit, or bills in Congress. Actions can have multiple types. An action may represent, say, the original creation of the item, a retweet, a comment, or a vote on the item. It is depicted by a typed edge between the corresponding user and claim. Action types are further associated with a valence of a known sign (i.e., positive or negative). For example, retweets (without modification) are generally positive-valence actions, whereas Reddit downvotes are negative valence actions.

\begin{figure}[!tb]
  \centering
  \includegraphics[width=0.47\textwidth]{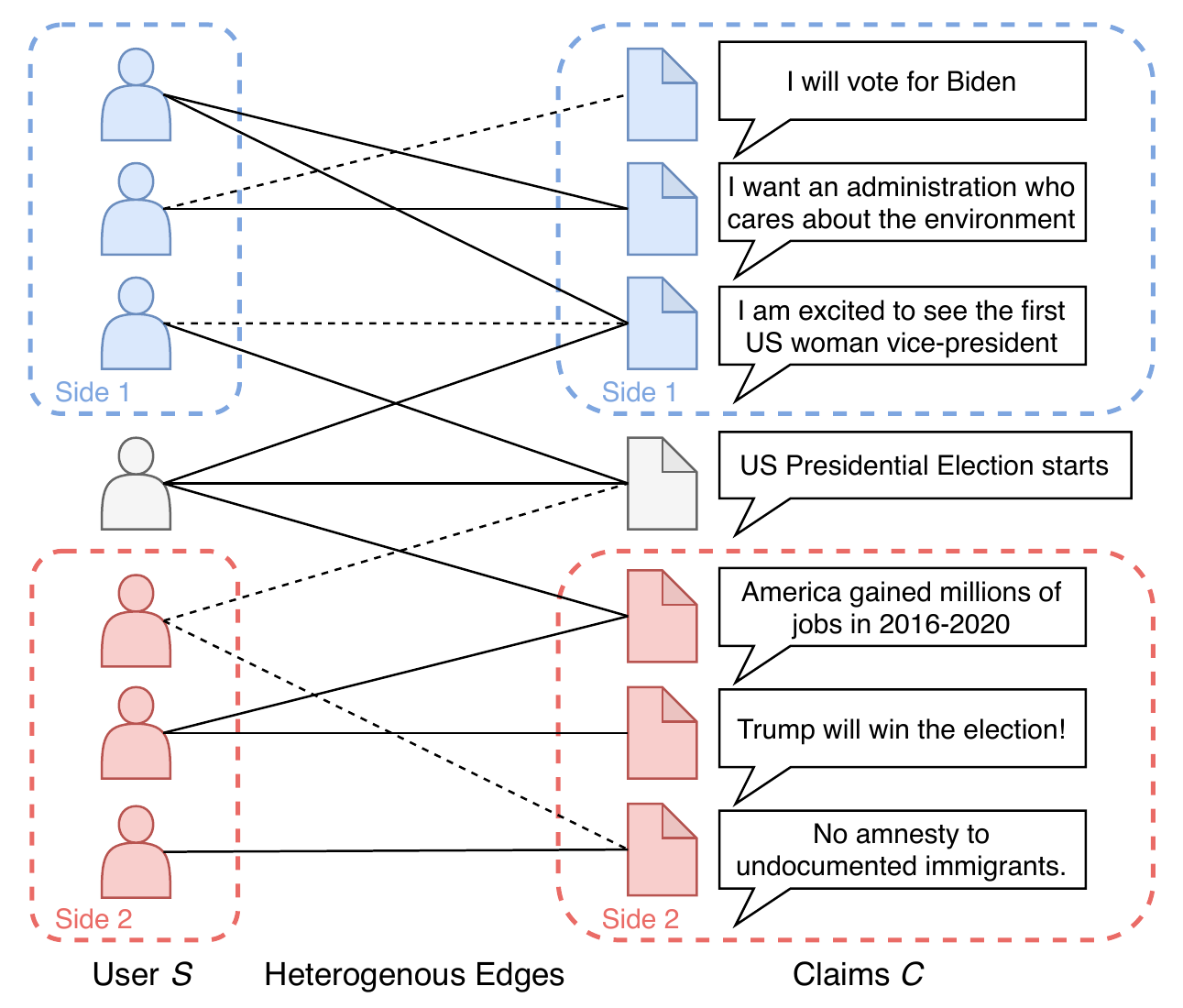}
  \caption{An example of users (left) and claims (right) from the US election, 2020. Solid and dashed edges represent two types of actions (for example, original posts and upvotes).}
  \label{fig:formulation}
  \vspace{-5pt}
\end{figure}

The objective of our algorithm is to find a latent representation, $z$, for users and claims that maximizes the likelihood of observed actions. We are particularly interested in polarized scenarios, where users and claims in the latent space are clustered, such that nearby users are more likely to take similar actions on similar claims, and nearby claims are more likely to be acted upon by similar users. Since ideologies unite groups by similar ideas, we expect users and items which share an ideology to be clustered in this latent space, without any supervision of polarity or ideology. As such, the latent representation, $z$ offers a basis for stance detection/prediction and ideology identification. Figure~\ref{fig:formulation} illustrates a simple example, where users make different claims that fall on two sides of a divide.


More formally, we model users and claims by a \textit{Bipartite Heterogeneous Information Network (BHIN)}~\cite{sun2012mining} given by a graph,  $\mathcal{G}=\{\mathcal{V}, \mathcal{E}\}$, where the number of vertices is $|\mathcal{V}|=N$ and the number of edges is $|\mathcal{E}|=M$. The number of vertex types is 
$2$. The number of edge types is
$R$. The BHIN can also be written as $\mathcal{G}=\{\{\mathcal{V}_1, \mathcal{V}_2\}, \{\mathcal{E}_1, \mathcal{E}_1, ..., \mathcal{E}_R\}\}$.
%
%
%
Each possible edge from the $i^{th}$ vertex to the $j^{th}$ vertex is denoted as $e_{ij}\in \mathcal{E}$ with a weight value $w_{ij}$. The dimensions of the adjacency matrix, $A$, are $(|\mathcal{V}_1| + |\mathcal{V}_2|) \times (|\mathcal{V}_1| + |\mathcal{V}_2|)$, where $A_{i,j} = w_{ij}$. We model $\mathcal{G}$ as an {\em undirected\/} graph, where $\left< v_i, v_j \right> \equiv \left< v_j, v_i \right>$.
Heterogeneity of edges allows expressing multiple types of actions. For instance, in our Congress example, different edge types represent the actions of voting \textit{Yea}, \textit{Nay}, or \textit{Abstain}. In general social networks, directionality is important (e.g., one might follow a celebrity but the celebrity might not follow back). Since our purpose, however, is to determine polarity, not social ties, we leverage the observation that action edges often imply the same relationship among polarities of endpoints, regardless of direction. For example, a ``star" action may be correlated with identical polarity whereas a ``downvote" action is more correlated with opposite polarity, regardless of the direction. Thus, the direction is ignored. 

The problem becomes converting an input BHIN into the (maximum likelihood) latent representation, $z$, for each user and claim.

\subsection{InfoVGAE: Belief Representation Learning}
\label{sec:model}

\begin{figure}[!ht]
  \centering
  \includegraphics[width=0.47\textwidth]{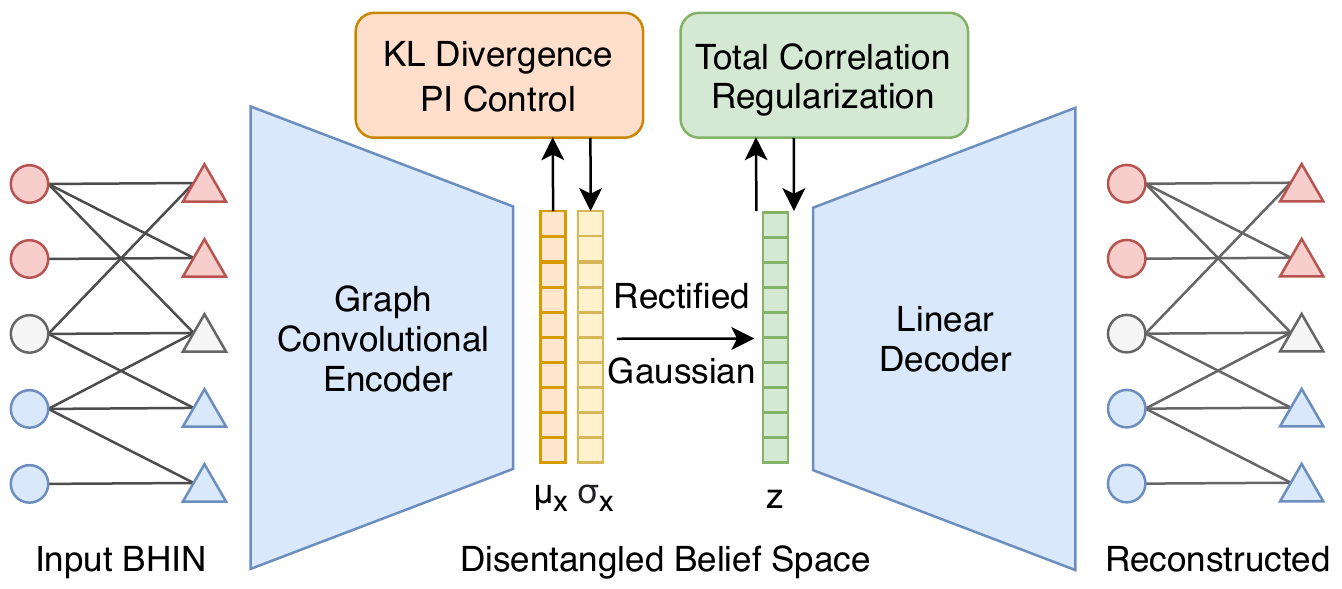}
  \caption{InfoVGAE consists of non-negative graph convolutional encoder, linear decoder, KL divergence control, and total correlation regularization. It encodes the user-claim bipartite graph into a latent space and then reconstructs it by sampling the data from the latent space distribution.}
  \label{fig::infovgae}
  \vspace{-5pt}
\end{figure}

We propose an Information-Theoretic Variational Graph Auto-Encoder (InfoVGAE) to map the users and claims into an explainable latent space for polarization analysis. The structure of InfoVGAE is shown in Figure~\ref{fig::infovgae}. Below, we describe the encoder and decoder, the learning process, and the use of InfoVGAE output by downstream tasks, such as stance or polarity detection.


\subsubsection{Non-Negative Inference Model (Encoder)}

The inference model takes a constructed BHIN as the input, denoted as $\mathcal{G} = \{\mathcal{V}, \mathcal{E}\}$. We use $\bm{A}$ to denote the adjacency matrix of $\mathcal{G}$ with self loops. $\bm{X}\in \mathbb{R}^{N\times F}$ is the input feature matrix.

We use $L$ layers Graph Convolutional Network (GCN)~\cite{kipf2016semi} as the network architecture of the encoder. Assume in the $l^{th}$ layer, the hidden state of GCN is $\bm{G}^{(l)}=GCN^{(l)}(\bm{A}, \bm{X})$, $\bm{G}^{(l)}\in \mathbb{R}^{N\times d_l}$, where $d_l$ is the dimension of hidden state in the $l^{th}$ layer. $\bm{G}^{(1)} = \bm{X}$ is the input feature matrix. $\bm{X}$ could be initialized as identity matrix $\bm{I}$ if there is no available feature. The GCN layer is formulated as
\begin{equation}
    \bm{G}^{(l)} = \gamma\left(
    \widetilde{\bm{A}}\bm{G}^{(l-1)}\bm{W}^{(l-1)}
    \right), ~ (2\leq l \leq L-1)
\end{equation}
where $\widetilde{\bm{A}} = \bm{D}^{-\frac{1}{2}}\bm{A}\bm{D}^{-\frac{1}{2}}$ is the symmetrically normalized adjacency matrix. $\bm{D}$ is diagonal degree matrix with the diagonal element $\bm{D}_{k, k} = \sum_{i=1}^{N} \bm{A}_{k_i}$. $\bm{W}_l\in \mathbb{R}^{d_l\times d_{l+1}}$ is the weight matrix in the $l^{th}$ layer. $\gamma$ denotes the activation function. We use Rectified Linear Unit (ReLU) as the activation function in our model. For the BHIN with multiple edge types ($R>1$), we extend GCN with Relational GCN~\cite{schlichtkrull2018modeling} for the encoder.

Assume $\bm{Z}\in \mathbb{R}^{N\times T}$ is the latent matrix. $T$ is the dimension of the target latent belief representation space. $\bm{z}_i$ is the latent vector of the $i^{th}$ node. The inference model is defined as:
\begin{equation}
    q(\bm{Z}|\bm{A}, \bm{X}) = \prod_{i=1}^{N} q(\bm{z}_i|\bm{A}, \bm{X}),~~~~q(\bm{z}_i|\bm{A}, \bm{X}) \sim \mathcal{N}_+(\bm{z}_i|\bm{\mu}_i, \bm{\sigma}_i^2),
\end{equation}
where posterior is assumed to follow the rectified Gaussian Distribution $\mathcal{N}_+=max(\mathcal{N}, 0)$~\cite{socci1998rectified}. $\bm{\mu}=\widetilde{\bm{A}}\bm{G}^{(L-1)}\bm{W}_{\bm{\mu}}^{(L-1)}$ is the matrix of mean vectors $\bm{\mu}_i$. $\log \bm{\sigma}=\widetilde{\bm{\bm{A}}}\bm{G}^{(L-1)}\bm{W}_{\bm{\sigma}}^{(L-1)}$ is the matrix of standard deviation vectors $\bm{\sigma}_i$. They share the hidden state $\bm{G}^{L-1}$.

The use of a rectified Gaussian Distribution ($\mathcal{N}_+$) to form the target latent space is not accidental. It stems from the observation we made in the introduction; namely, while the positions espoused by opposing ideologies are generally in conflict, they are not necessarily {\em opposite\/} and, for that matter, not always mutually exclusive. For example, some might favor funding the military, while others might favor funding environmental research. In fact, some might believe in doing both. This is in contrast to, say, simple {\em stance detection\/}, where stance on some topic is either positive, neutral, or negative. Thus, in representing {\em systems of belief\/}, we remove the negative side of each latent axis, thereby forcing the disentanglement of different ideologies onto {\em different axes\/} (as opposed to a single axis with positive versus negative values). This representation offers better compositionality. For example, individuals who mix and match elements of different belief systems (e.g., strongly believe in funding both the military and the environment) can now be better represented.
Polarization is more prominent when these individuals and content align more closely with individual axes in our latent space (and away from the origin) as opposed to being closer to the diagonals. We see in the evaluation section that our non-negative target space indeed helps separate polarized nodes into corresponding axes.

\begin{figure*}[!ht]
  \centering
\begin{subfigure}[b]{0.139\textwidth}
    \includegraphics[width=\textwidth]{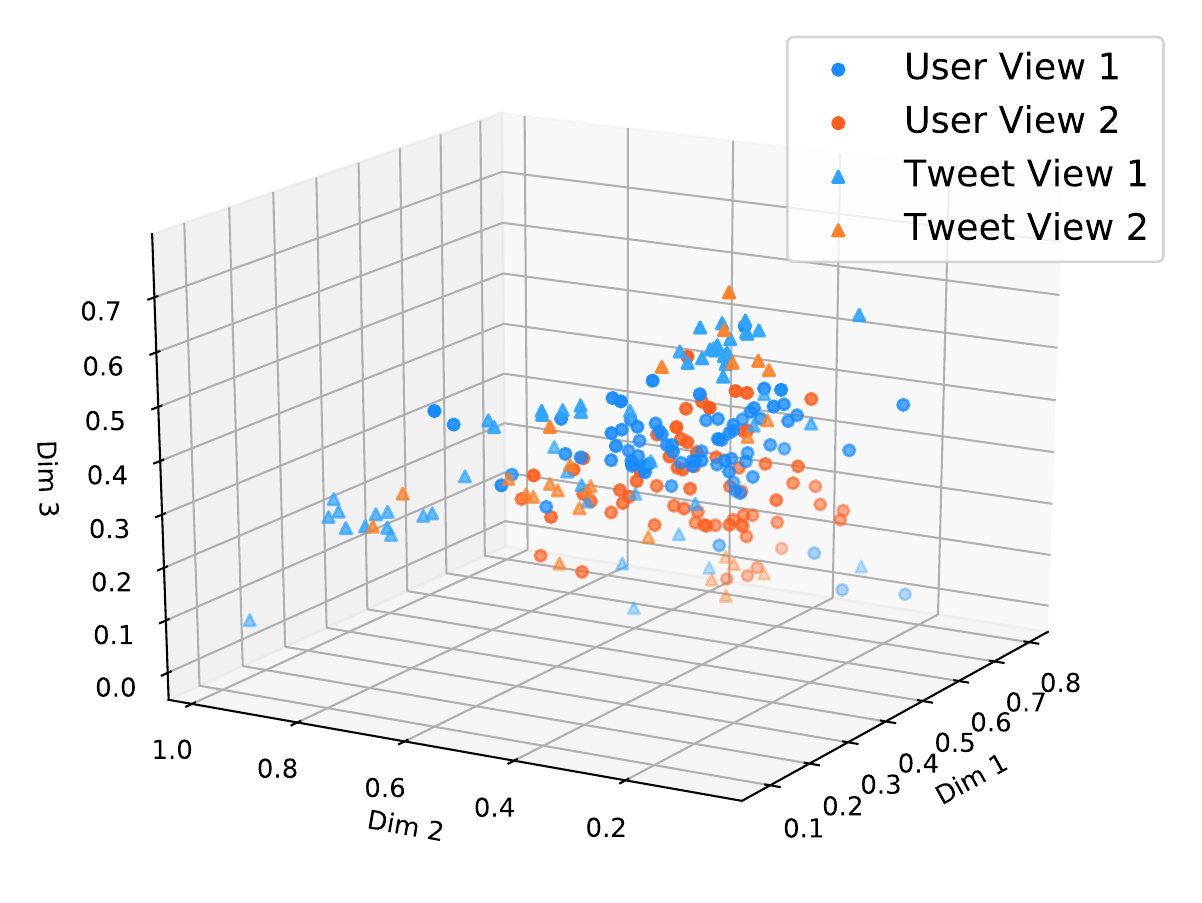}
\end{subfigure}
\begin{subfigure}[b]{0.139\textwidth}
    \includegraphics[width=\textwidth]{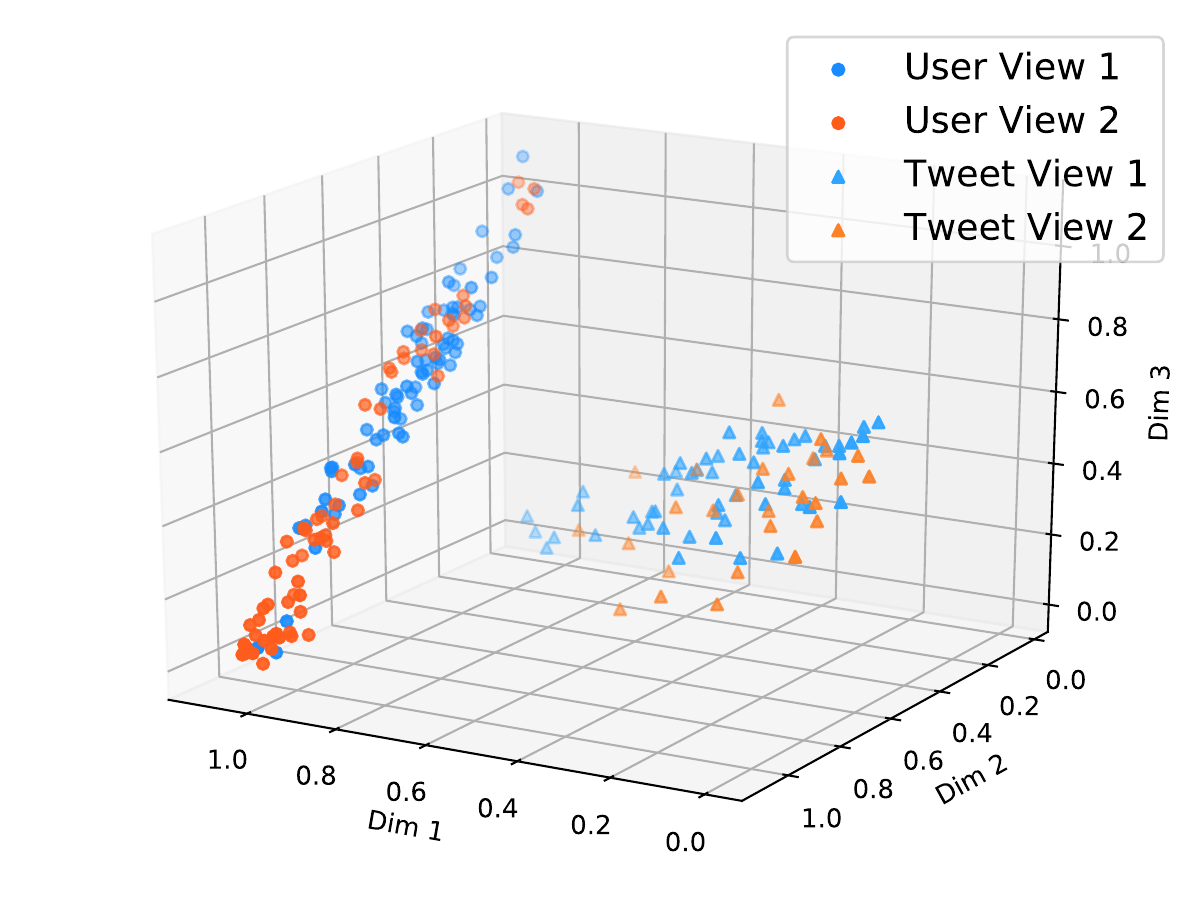}
\end{subfigure}
\begin{subfigure}[b]{0.139\textwidth}
    \includegraphics[width=\textwidth]{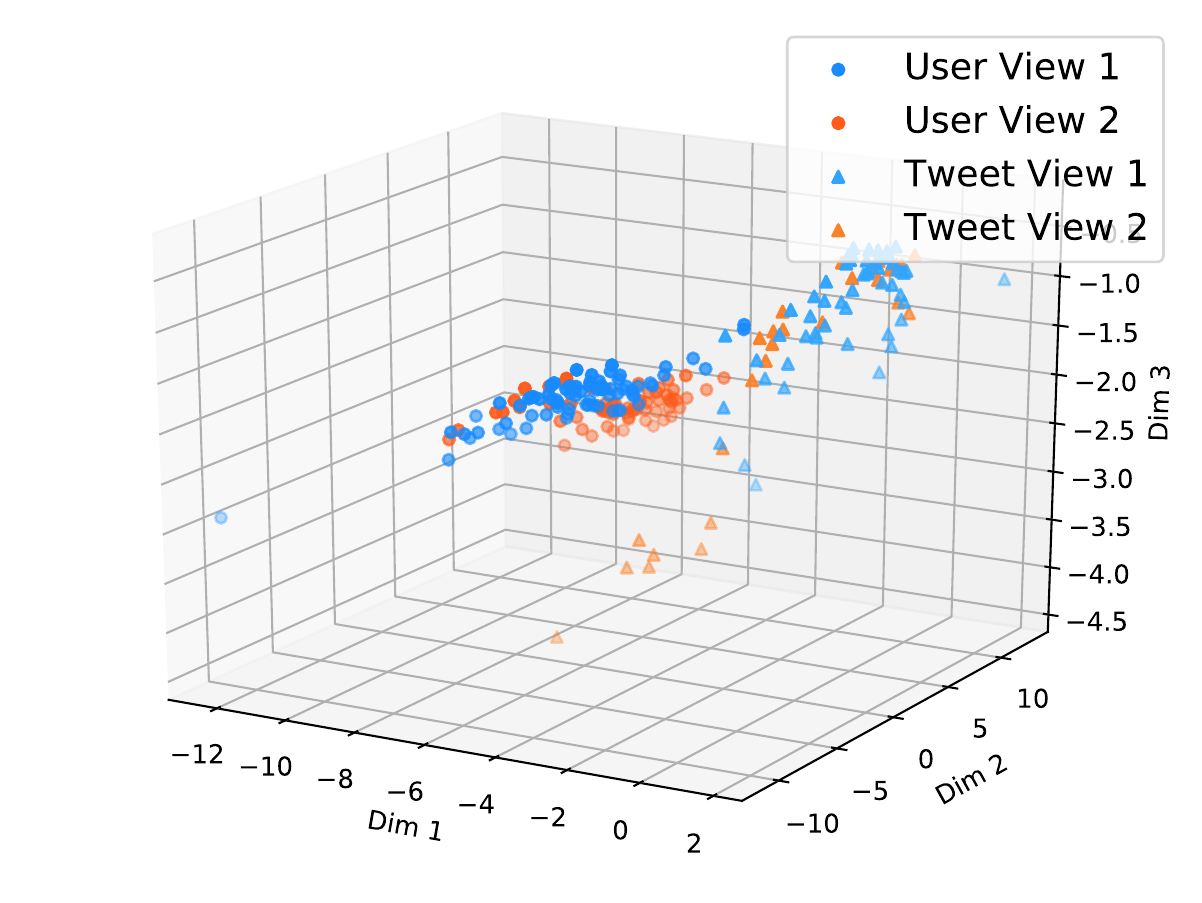}
\end{subfigure}
\begin{subfigure}[b]{0.139\textwidth}
    \includegraphics[width=\textwidth]{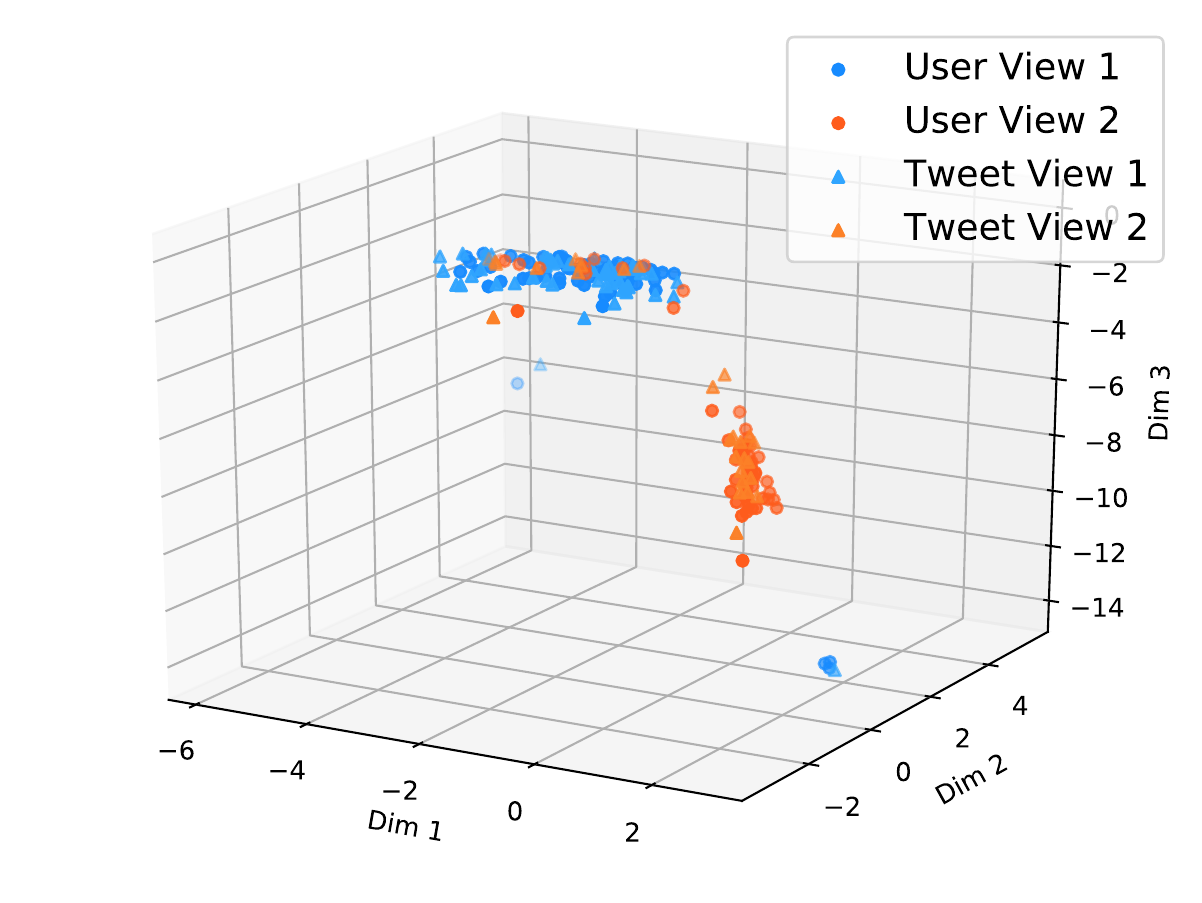}
\end{subfigure}
\begin{subfigure}[b]{0.139\textwidth}
    \includegraphics[width=\textwidth]{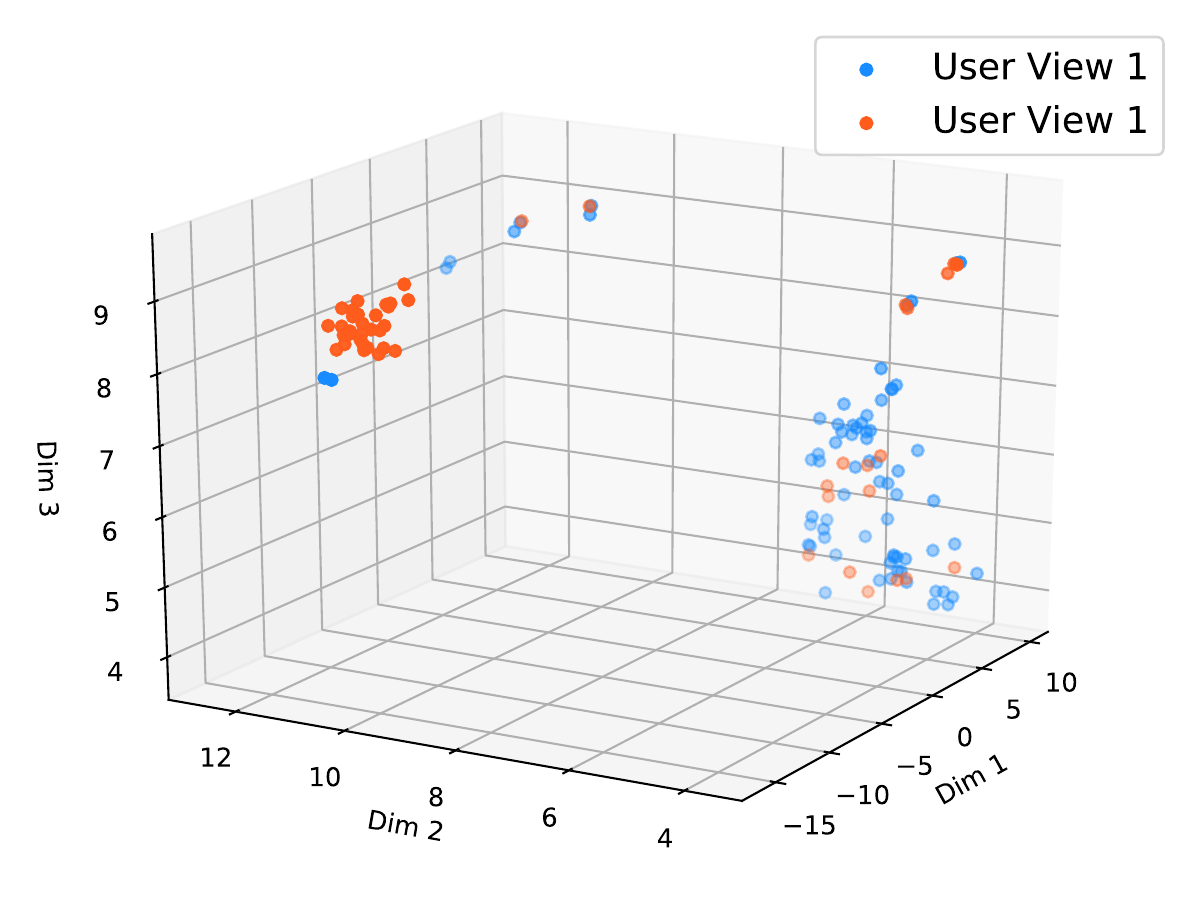}
\end{subfigure}
\begin{subfigure}[b]{0.139\textwidth}
    \includegraphics[width=\textwidth]{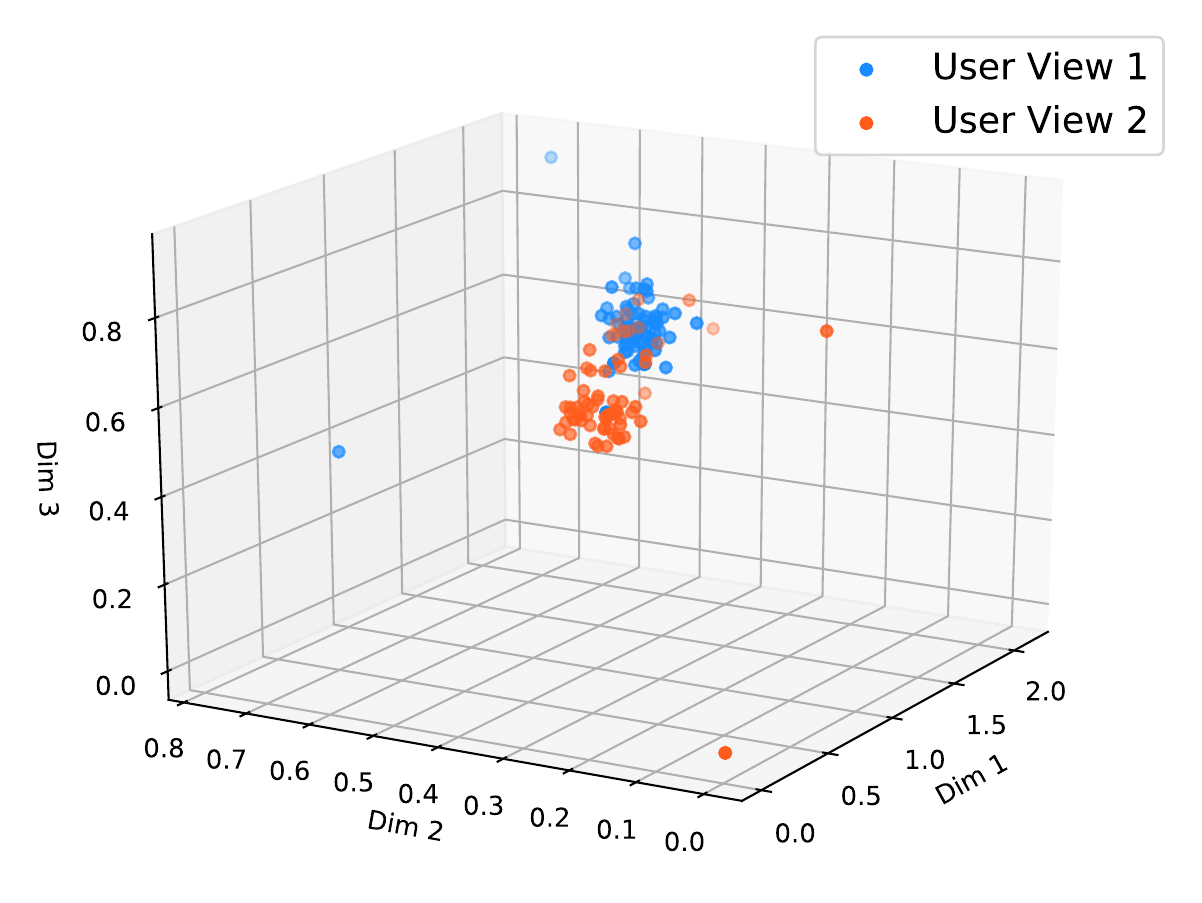}
\end{subfigure}
\begin{subfigure}[b]{0.139\textwidth}
    \includegraphics[width=\textwidth]{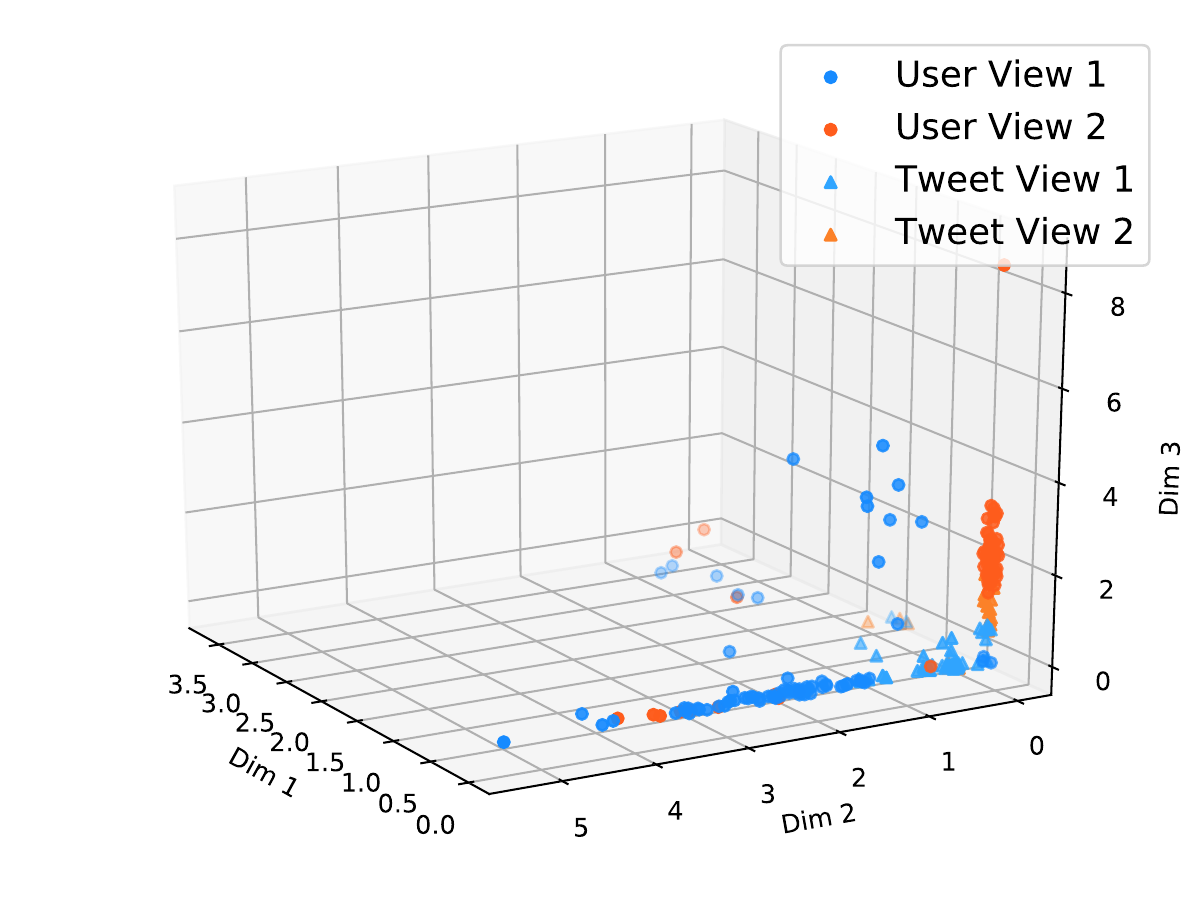}
\end{subfigure}\\
\begin{subfigure}[b]{0.139\textwidth}
    \includegraphics[width=\textwidth]{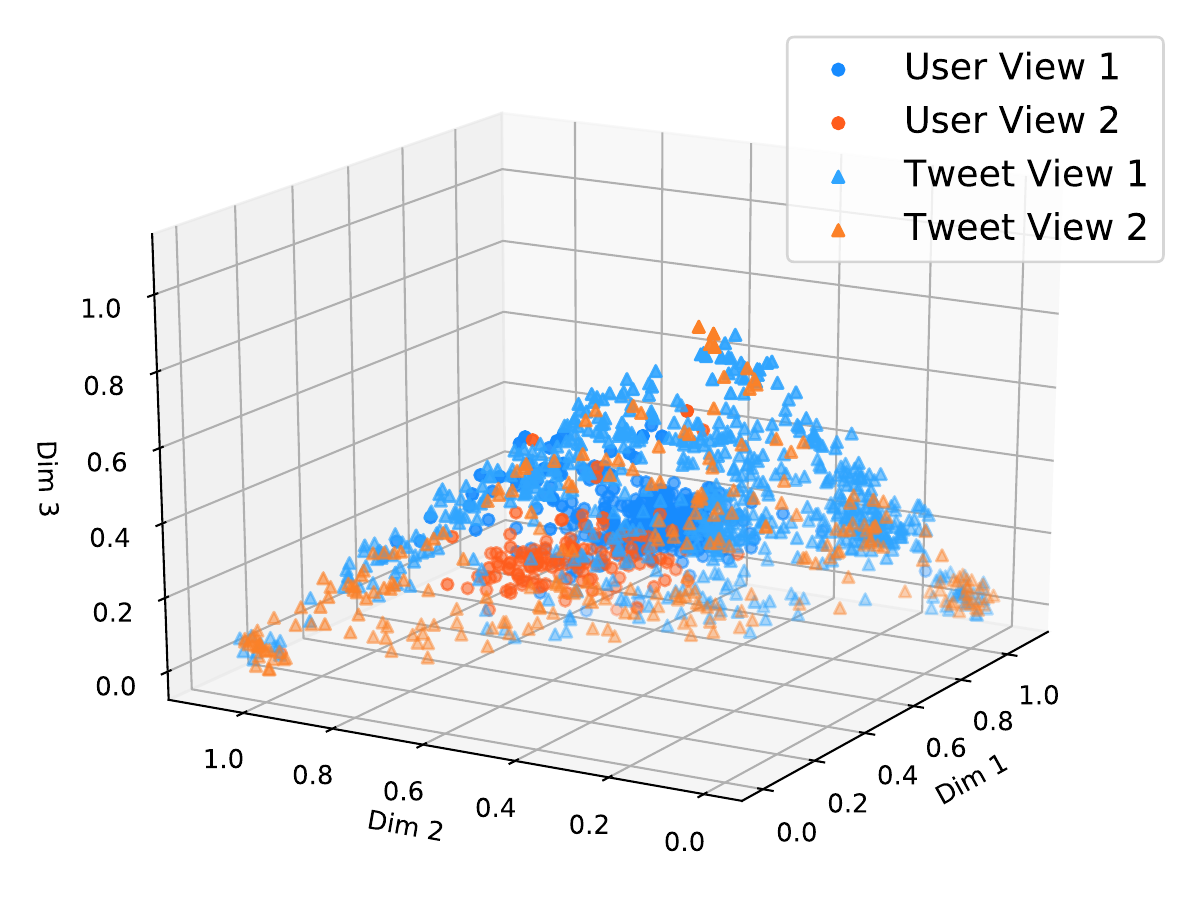}
\end{subfigure}
\begin{subfigure}[b]{0.139\textwidth}
    \includegraphics[width=\textwidth]{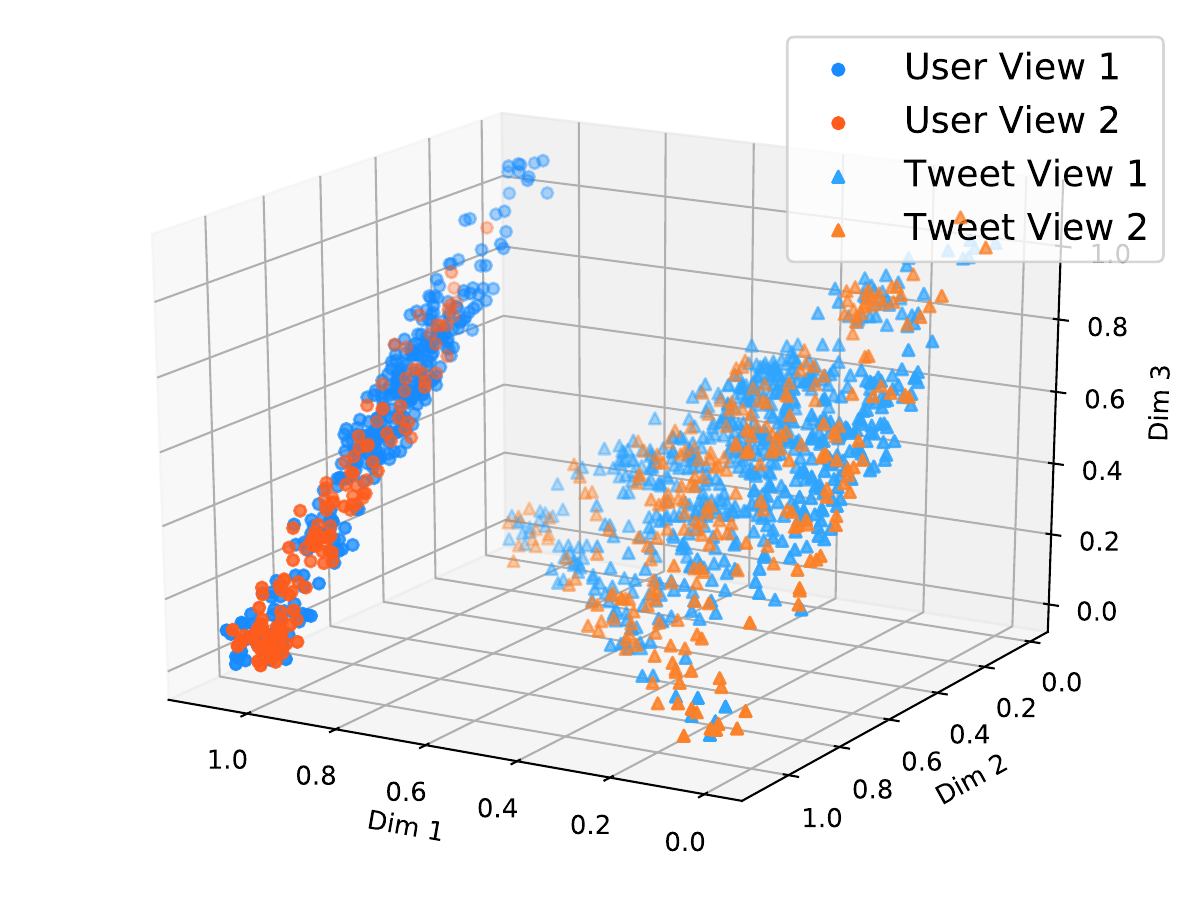}
\end{subfigure}
\begin{subfigure}[b]{0.139\textwidth}
    \includegraphics[width=\textwidth]{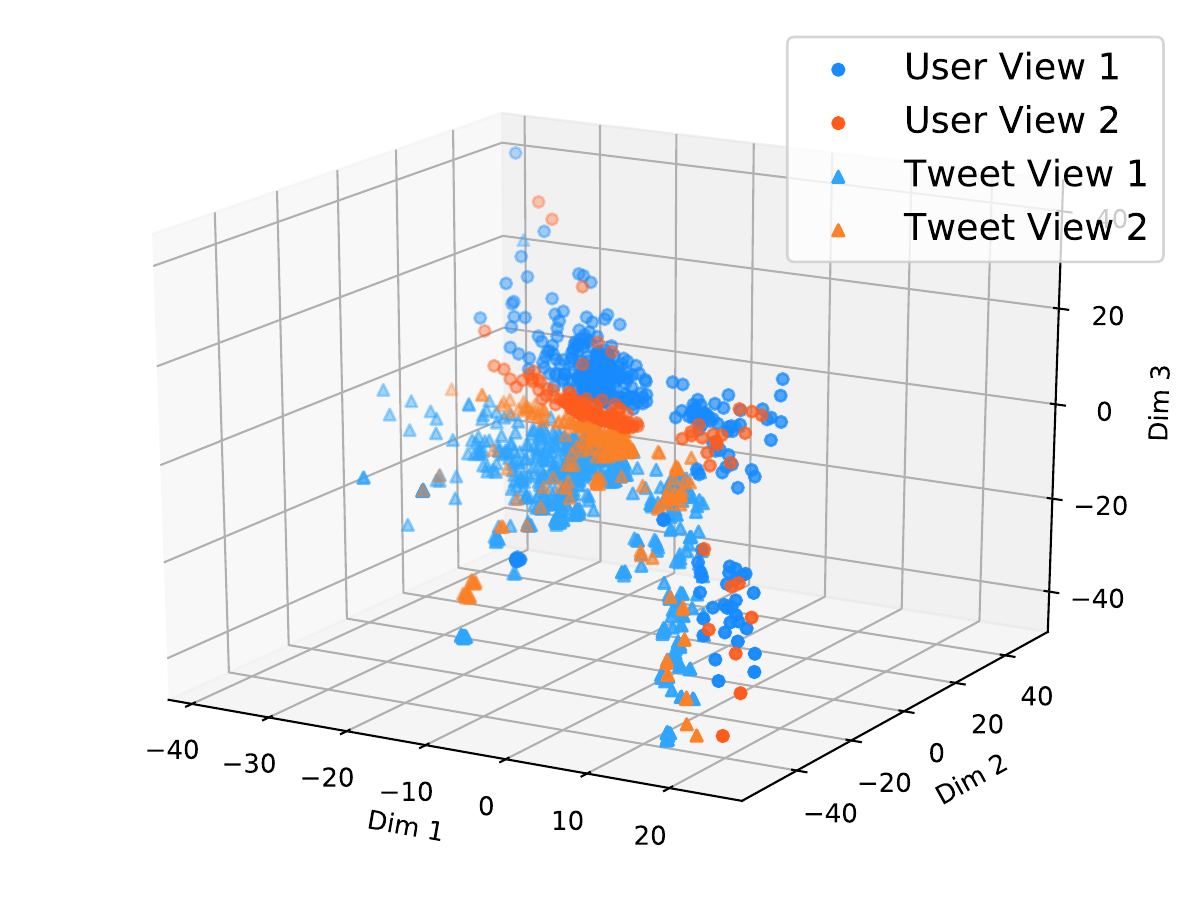}
\end{subfigure}
\begin{subfigure}[b]{0.139\textwidth}
    \includegraphics[width=\textwidth]{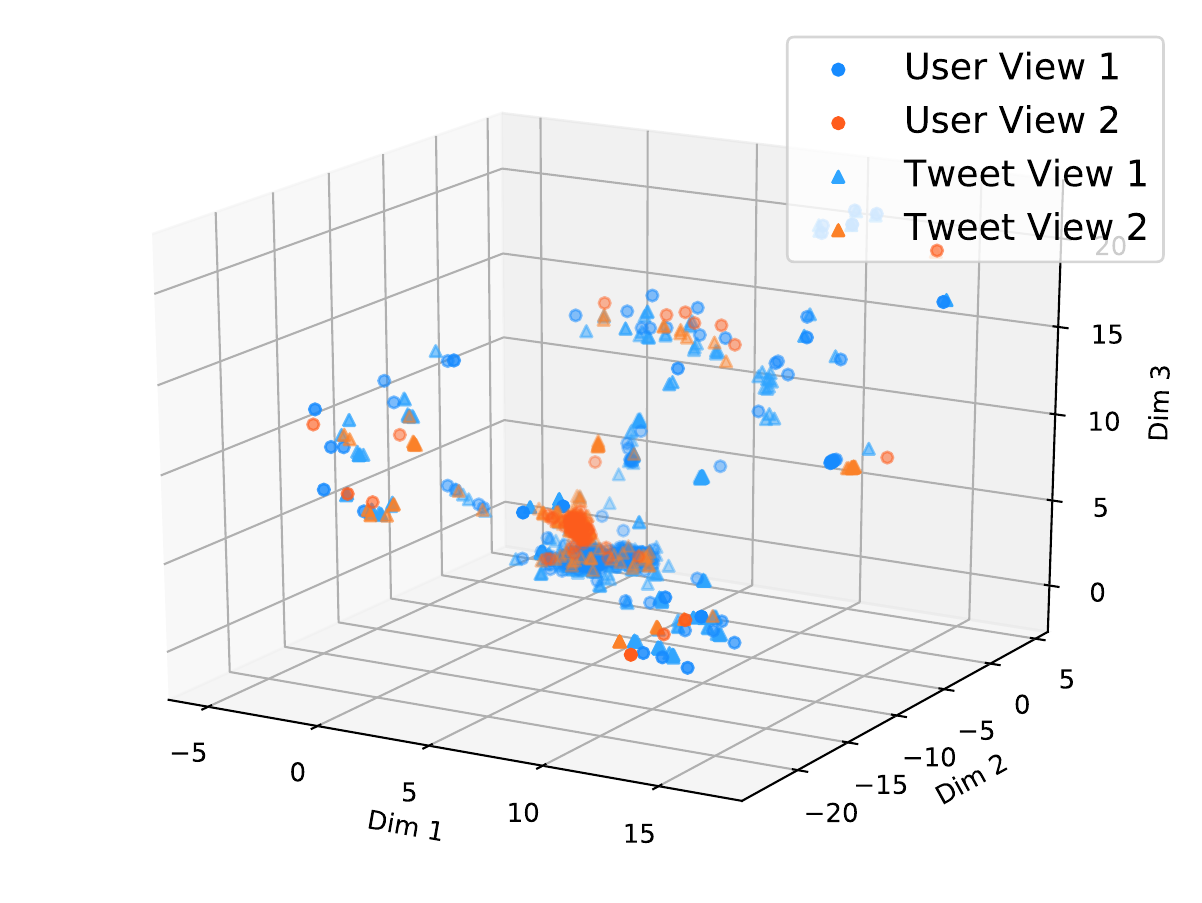}
\end{subfigure}
\begin{subfigure}[b]{0.139\textwidth}
    \includegraphics[width=\textwidth]{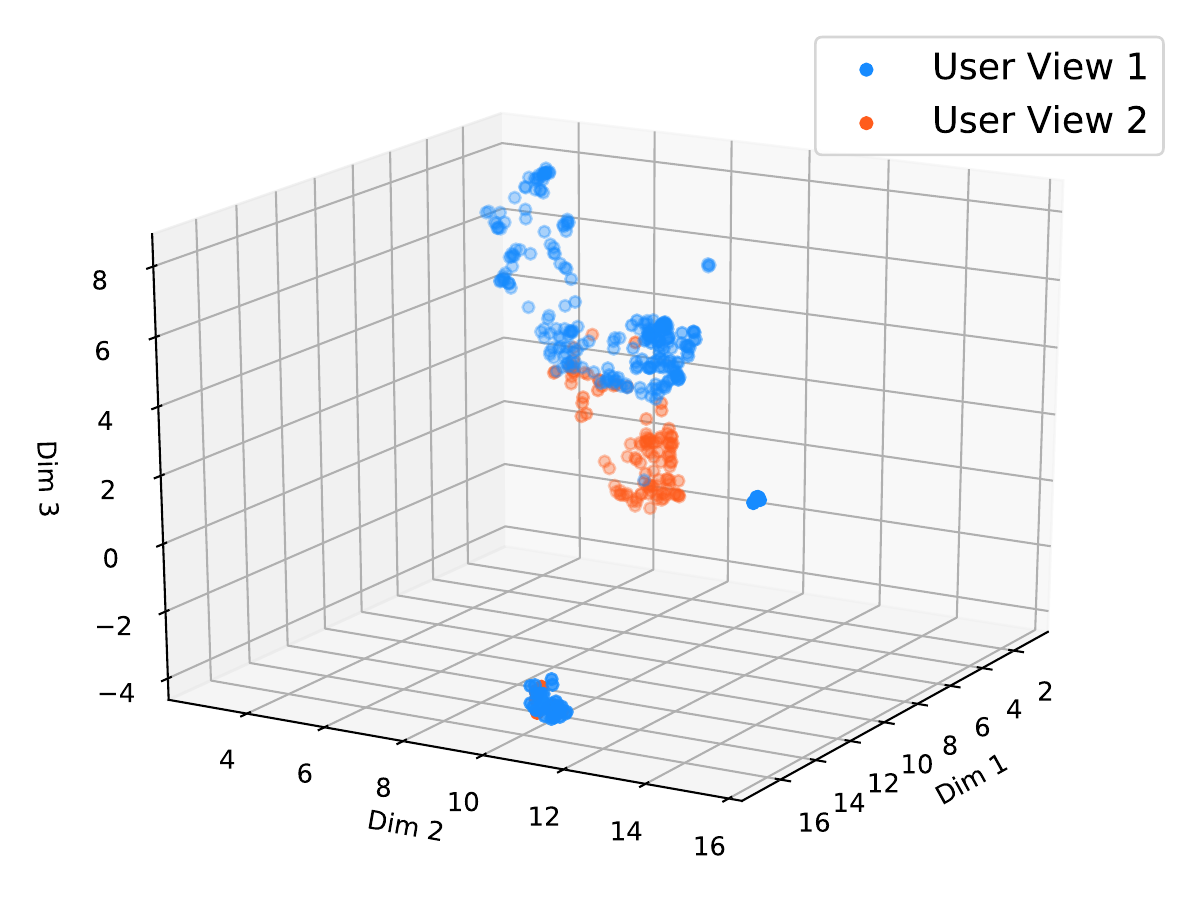}
\end{subfigure}
\begin{subfigure}[b]{0.139\textwidth}
    \includegraphics[width=\textwidth]{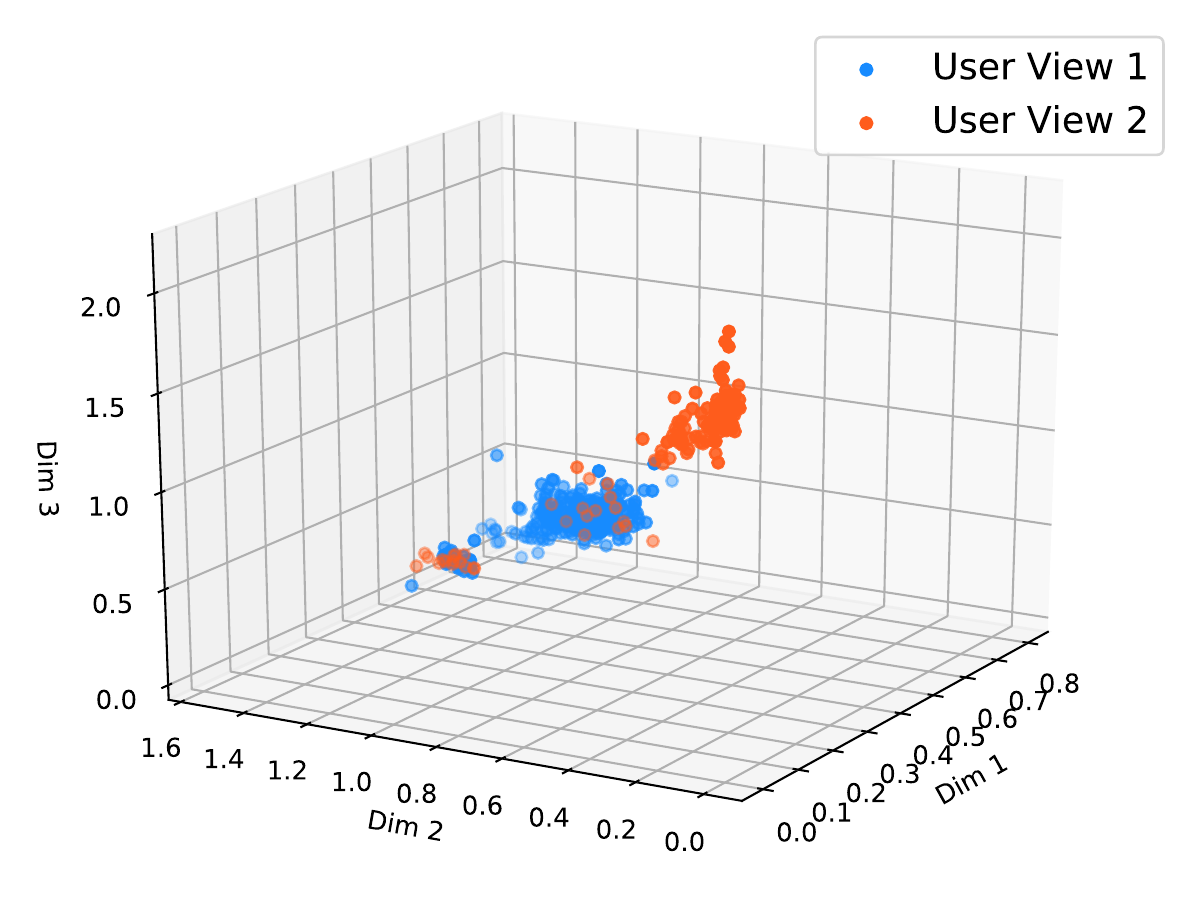}
\end{subfigure}
\begin{subfigure}[b]{0.139\textwidth}
    \includegraphics[width=\textwidth]{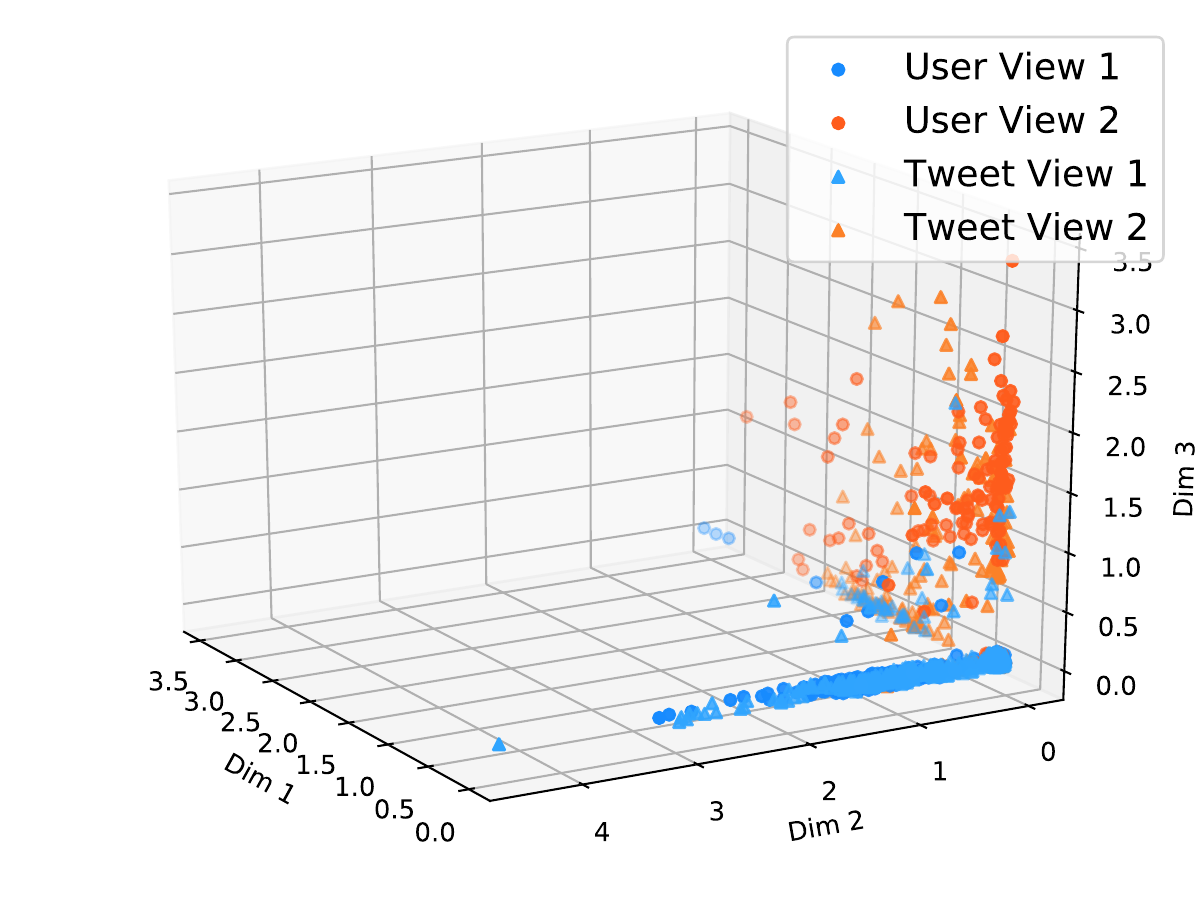}
\end{subfigure}\\
\begin{subfigure}[b]{0.139\textwidth}
    \includegraphics[width=\textwidth]{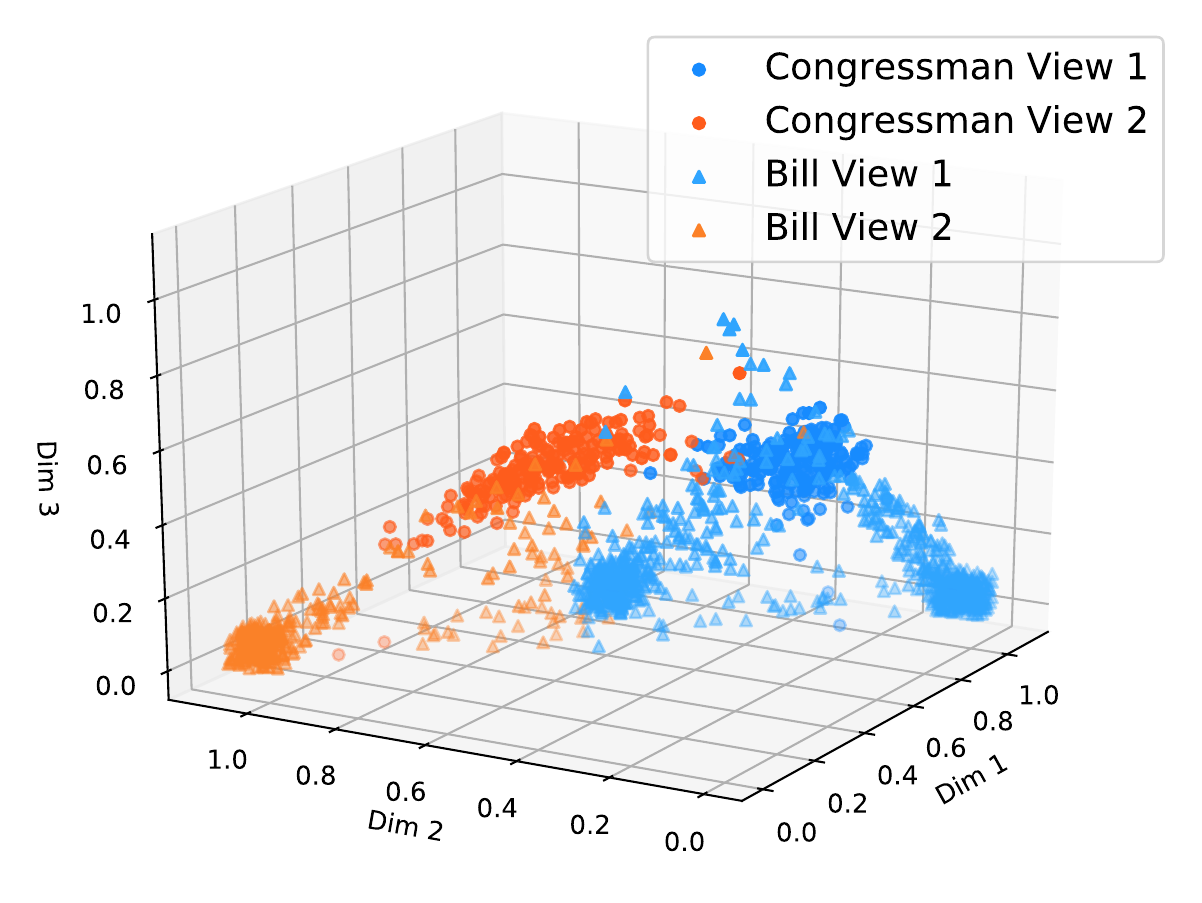}
\caption{NMF}
\label{fig::embeddingNMF}
\end{subfigure}
\begin{subfigure}[b]{0.139\textwidth}
    \includegraphics[width=\textwidth]{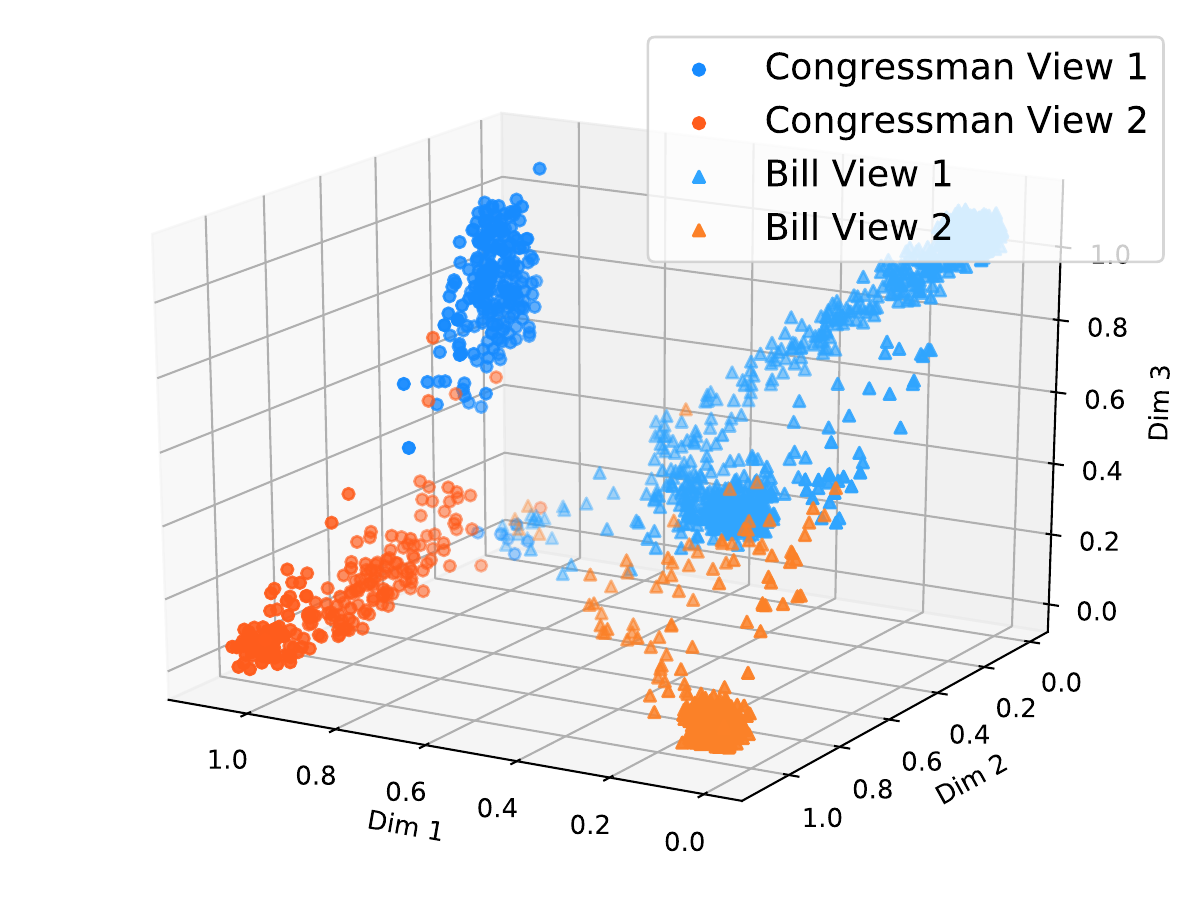}
\caption{BSMF}
\label{fig::embeddingBSMF}
\end{subfigure}
\begin{subfigure}[b]{0.139\textwidth}
    \includegraphics[width=\textwidth]{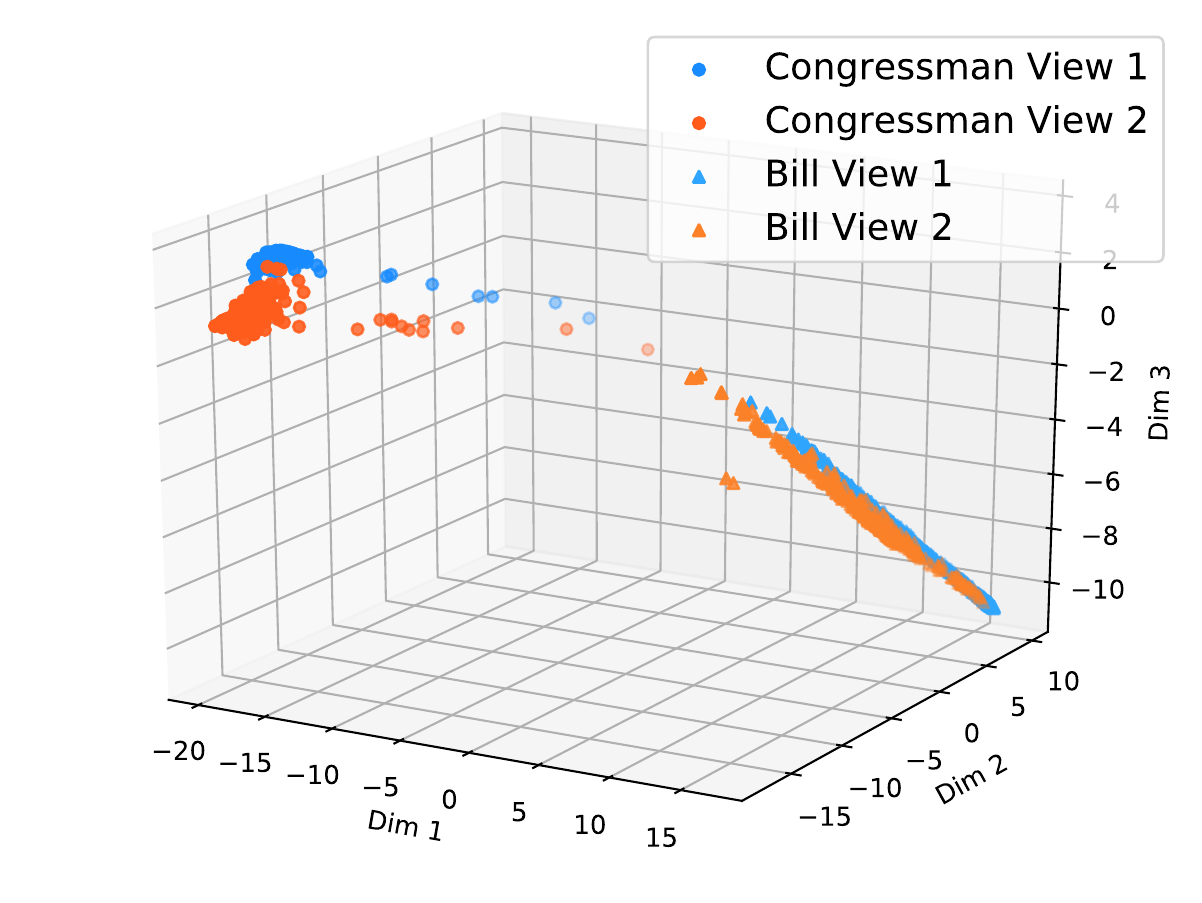}
\caption{GCN}
\label{fig::embeddingGCN}
\end{subfigure}
\begin{subfigure}[b]{0.139\textwidth}
    \includegraphics[width=\textwidth]{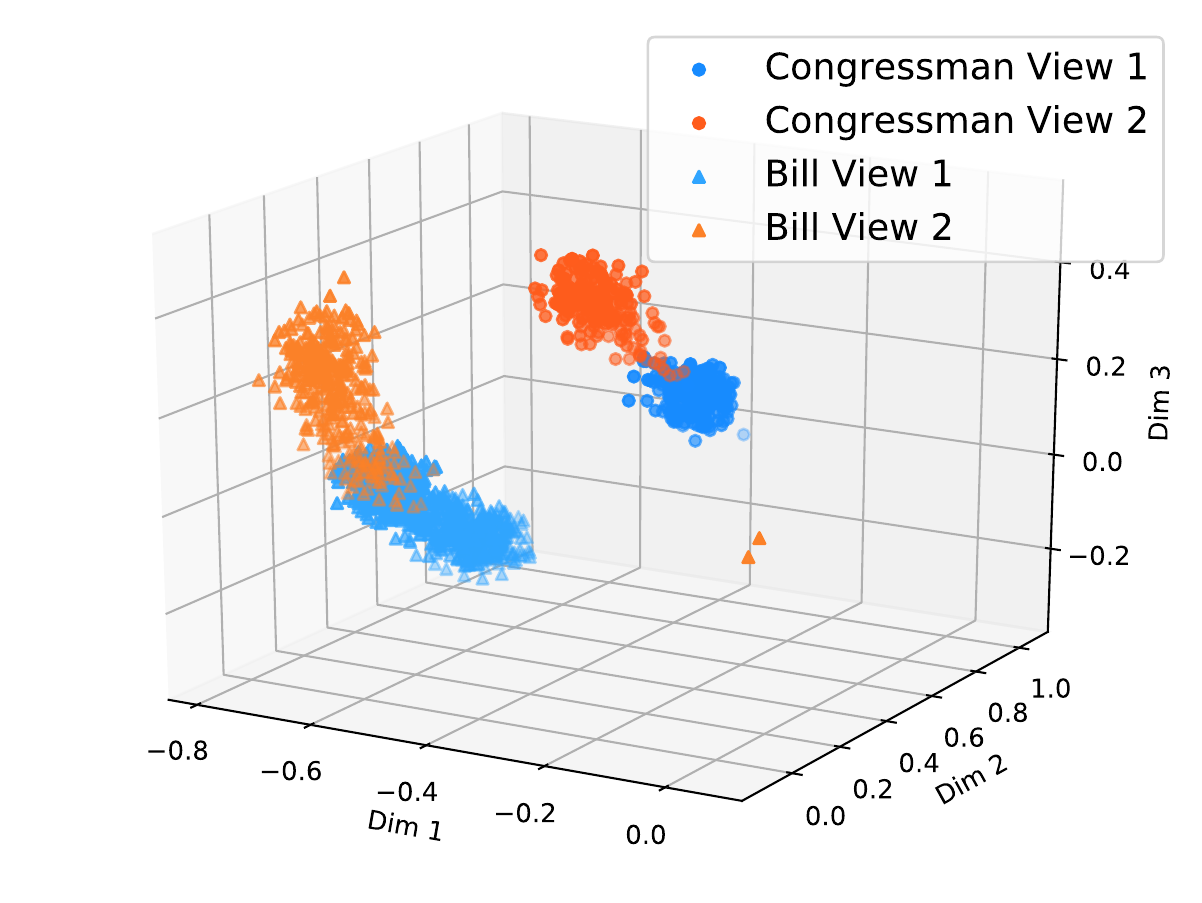}
\caption{DeepWalk}
\label{fig::embeddingDeepWalk}
\end{subfigure}
\begin{subfigure}[b]{0.139\textwidth}
    \includegraphics[width=\textwidth]{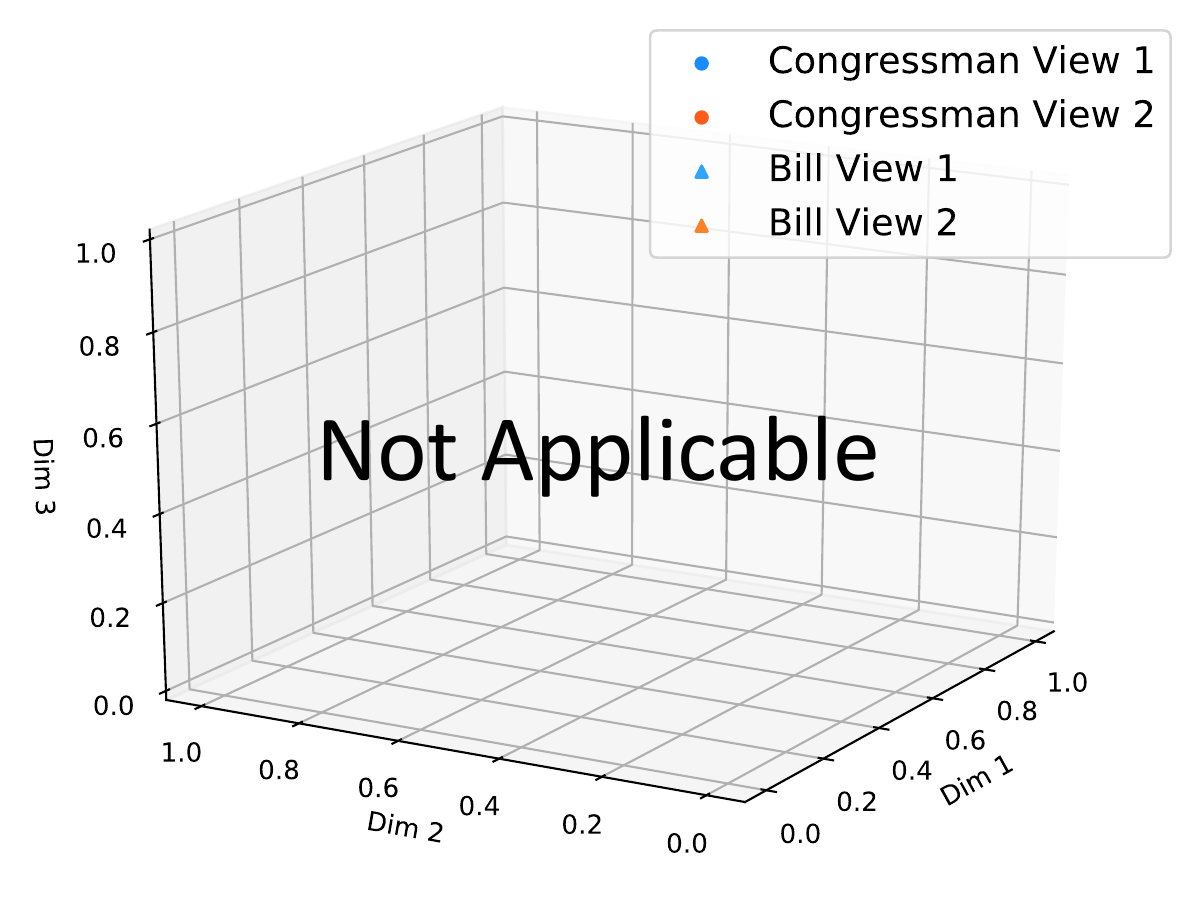}
\caption{Stance}
\label{fig::embeddingStance}
\end{subfigure}
\begin{subfigure}[b]{0.139\textwidth}
    \includegraphics[width=\textwidth]{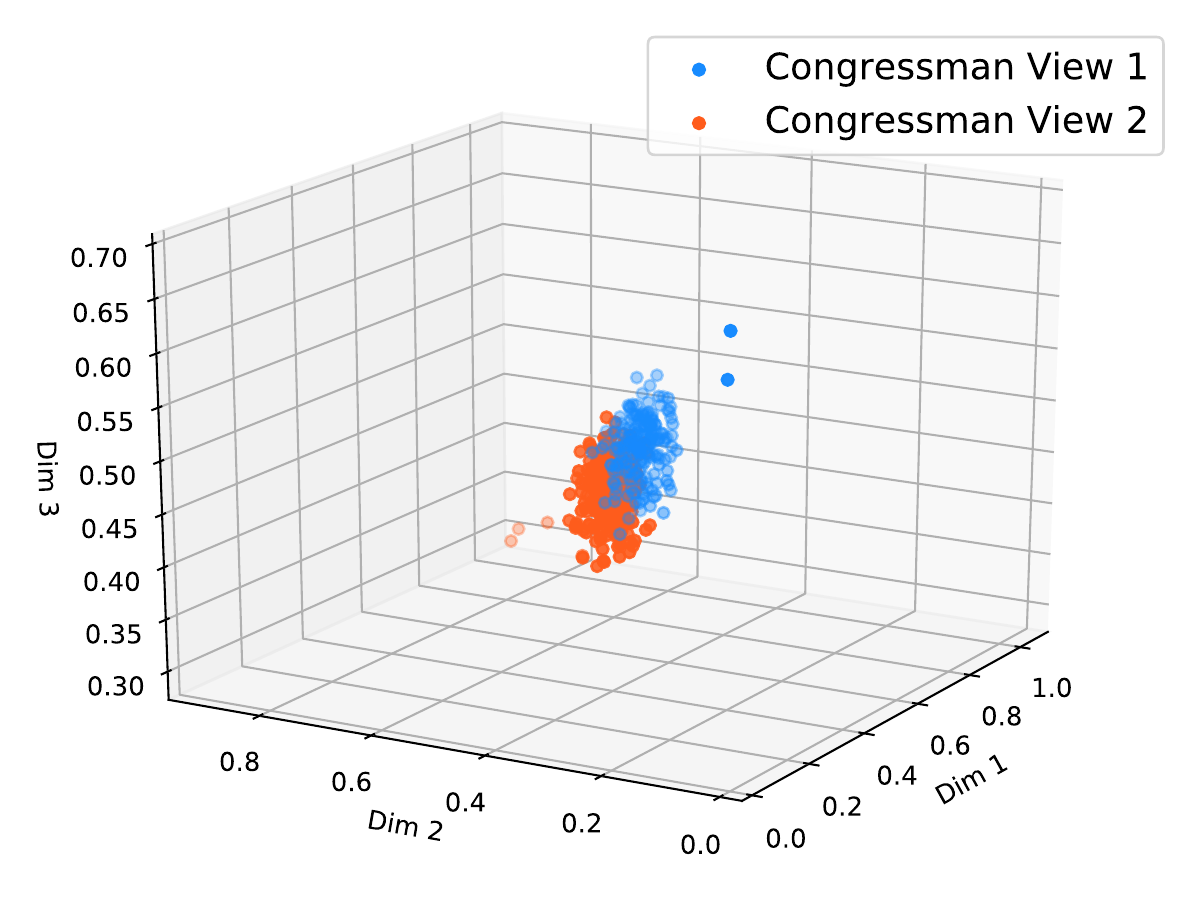}
\caption{TIMME-Unsup}
\label{fig::TIMME}
\end{subfigure}
\begin{subfigure}[b]{0.139\textwidth}
    \includegraphics[width=\textwidth]{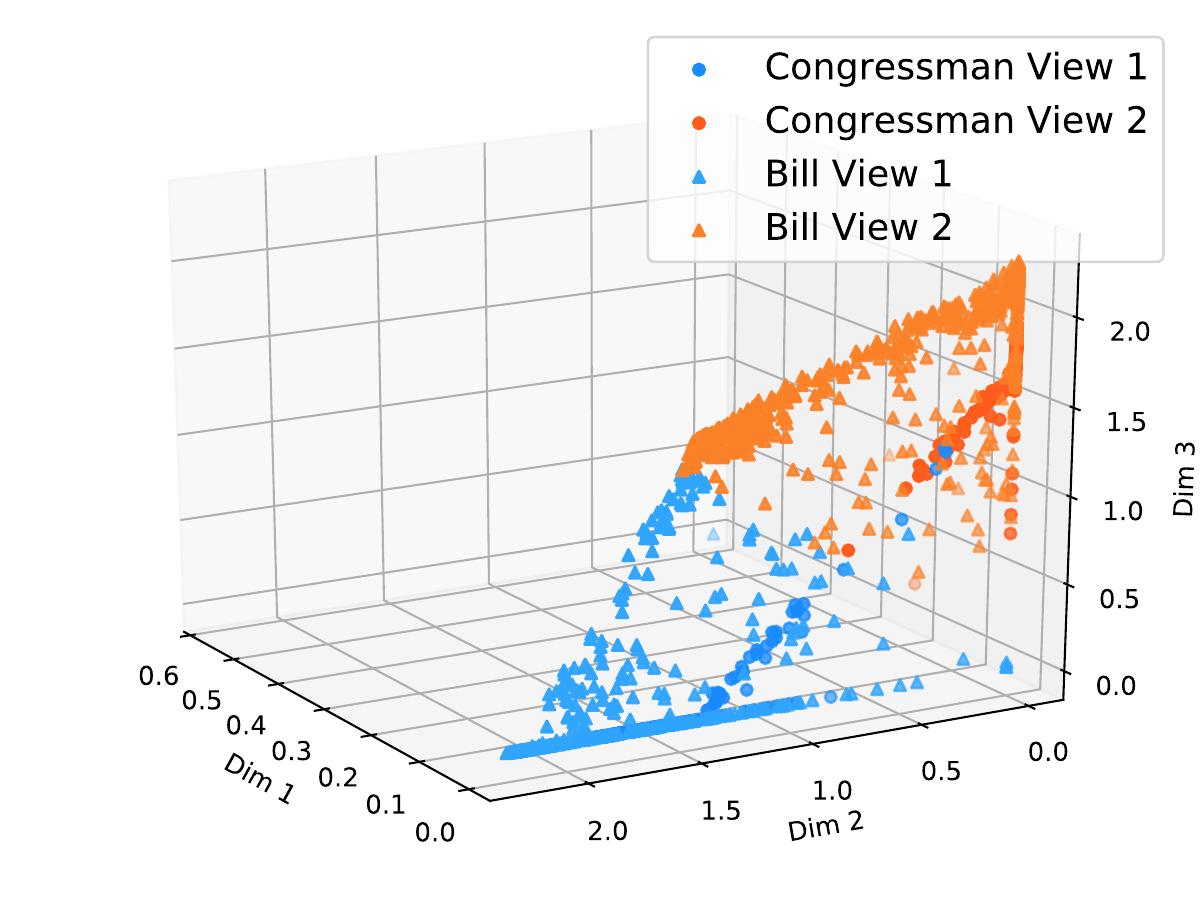}
\caption{InfoVGAE(Ours)}
\label{fig::embeddingOUR}
\end{subfigure}
  \caption{Representations of users and claims in latent space. The results on US Election, Eurovision, and Voteview datasets are shown in the first, second, and third rows, respectively. The data points are colored according to labels. Only labeled points are plotted. Ideal representations should separate the points of different colors clearly. Note that Stance Detection (Stance) and Unsupervised TIMME (TIMME-Unsup) methods in Figure\ref{fig::embeddingStance} and Figure\ref{fig::TIMME} can only produce the embedding of users. Stance Detection method is not applicable to Voteview dataset, because it requires Twitter platform-specific data such as hashtag.}
  \label{fig::embedding}
  \vspace{-12pt}
\end{figure*}

\subsubsection{Generative Model (Decoder)}
We use a linear inner product decoder. Its generative model can be formulated as:
\begin{equation}
    p(\bm{A} | \bm{Z}) = \prod_{i=1}^N \prod_{j=1}^{N}
    p(\bm{A}_{i,j} | \bm{z}_i, \bm{z}_j),~~~~
    p(\bm{A}_{i,j} | \bm{z}_i, \bm{z}_j) = \sigma(\bm{z}_i^T\bm{z}_j),
\end{equation}
where $\sigma(\cdot)$ is the logistic sigmoid function. This inner product decoder enhances the explainability of our latent space. In the geometric space, $\bm{z}_i^T\bm{z}_j$ is defined as the multiplication between the norm of the projection of $\bm{z}_i$ over $\bm{z}_j$ and the norm of $\bm{z}_j$. While we maximize $p(\bm{A}_{i,j} | \bm{z}_i, \bm{z}_j)$, we are forcing the latent vectors $\bm{z}_i, \bm{z}_j$ to be parallel if $\bm{A}_{i,j} = 1$ and orthogonal otherwise. As a result, different systems of belief will be associated with different orthogonal axes.

\subsubsection{Total Correlation Regularization}\label{sec::total}
Inspired by Total Correlation (TC)~\cite{watanabe1960information} in information theory, we design a total correlation regularization module in InfoVGAE to encourage the model to disentangle the latent variables. The total correlation is a generalization of mutual information for multiple random variables. By penalizing the total correlation term, the model is able to learn a series of statistically independent latent variables, thereby making the belief representations interpretable.

Let $z_i^k$ denote the $k^{th}$ dimension of the latent vector for the $i^{th}$ node, the total correlation regularizer is defined as
\begin{align}
    \mathcal{L}_{TC}(\bm{z}_i) = D_{KL}[ q(\bm{z}_i) \| \prod_{k=1}^{d} q(\bm{z}_i^k)) ],
\end{align}
which can be interpreted as the divergence between joint distribution and independent distribution. In our case, however, this KL divergence is intractable, because the entire dataset is required to evaluate every $q(\bm{z}_i)$ with direct Monte Carlo estimate. There are two main approaches in the literature to tackle this problem: (i) Decompose the KL divergence and then use Monte Carlo sampling to compute the independent probability~\cite{chen2019isolating,zhao2019infovae,gao2019auto}; (ii) Train a discriminator to approximate the KL divergence with density-ratio trick~\cite{nguyen2010estimating,kim2018disentangling,sugiyama2012density}. In this paper, we train a discriminator $\Phi(\bm{z}_i)$ to discriminate whether a sample $\bm{z}_i$ is from the joint distribution $q(\bm{z}_i)$ or independent distribution $\prod_{k=1}^{d}q(\bm{z}_i^k)$. The total correlation thus can be approximated by
\begin{equation}
    \mathcal{L}_{TC}(\bm{z}_i) \approx \mathbb{E}_{\bm{z}_i\sim q(\bm{z}_i)}[\log(\Phi(\bm{z}_i)) - \log(1 - \Phi(\bm{z}_i))].
\end{equation}
To jointly train the discriminator and VGAE model, in each training step of VAE, we sample a batch of $z_i$ from $q(z_i)$ as positive samples (joint distribution), and generate the negative samples (independent distribution) with the independence trick~\cite{arcones1992bootstrap}. For every latent dimension, we randomly permute the values across different items in the batch. Then the parameters of discriminator can be optimized via the maximum likelihood estimation.

\subsubsection{Joint Training with KL Divergence PI Control Module}
The design of VAEs often suffers from KL-vanishing, also called posterior collapse, in that the value of KL-divergence becomes zero during model training. This implies over-fitting and entails a failure to generalize from training data. This phenomenon is especially significant for graph representation learning models~\cite{sun2021interpretable}. We introduce the Proportional Integral (PI) control module to tackle the KL vanishing problem and ensure a disentangled latent space. The PI controller~\cite{shao2020controlvae} can dynamically tunes $\beta(t)$ to manipulate the KL-divergence based on the difference between the actual value and the target value during model training. The control process can be formulated as:
\begin{equation}\label{eq:PI}
\beta(t) = \frac{K_p}{1+\exp(e(t))} - K_i \sum_{j=0}^t e(j),
\end{equation}
where $e(t)$ is the difference between the target $KL_{set}$ and the actual KL-divergence at training step, $t$. $K_p$ and $K_i$ are positive hyper-parameters of the designed PI controller. The error pushes $\beta(t)$ in a direction that causes KL-divergence to approach the target. When the KL-divergence is too small, $e(t)$ becomes positive, causing the output, $\beta(t)$, of the PI controller to decrease, thereby boosting the actual KL-divergence to higher values. This mechanism encourages InfoVGAE to learn a more informative latent representation.

The overall objective of InfoVGAE includes optimizing the evidence lower bound (ELBO) of VAE while simultaneously minimizing the total correlation, which can be formulated as
\begin{align}
\begin{aligned}
\mathbb{E}_{\bm{Z}\sim q(\bm{Z}|\bm{A}, \bm{X})} &\left[\log p(\bm{A} | \bm{Z})\right] - \beta(t) D_{KL}\left[q(\bm{Z}|\bm{A}, \bm{X}) \| p(\bm{Z})\right]\\
&\hspace{8pt} - \lambda \mathbb{E}_{\bm{Z}\sim q(\bm{Z})}[\log(\Phi(\bm{Z})) - \log(1 - \Phi(\bm{Z}))],
\end{aligned}
\end{align}
where the first two terms are the ELBO objective with PI control variable $\beta(t)$. The last term is the total correlation regularizer introduced in Section~\ref{sec::total}. The joint training process of VAE, total correlation regularizer, and PI control module brings additional benefits for InfoVGAE to learn a disentangled and informative latent representation.

\subsubsection{Downstream Tasks}\label{sec::unsupervised}
Once a latent representation is learned, several downstream tasks become possible. The easiest is to determine the stance or ideological polarity espoused by users and claims (depending on whether the input data comprises opinions on one topic, as in stance, or views on a number of different topics, thus revealing a system of beliefs). The disentangled latent space produced by the InfoVGAE offers a simpler way to separate such stances or polarities. As is shown in Figure~\ref{fig::embeddingOUR}, every axis is associated with a different latent belief system. To select the dominant ones, we choose the axes with the largest accumulated projection values over all the data points. We then use point coordinates along those axes as measures of alignment with the corresponding ideologies. Thus, we can classify the polarity of a user or claim simply by considering the axis where it has the largest coordinate (without using a clustering algorithm). We can also predict the likelihood that a user agrees with a claim from their proximity in the latent space. We can also rank users and claims on each axis by the corresponding coordinate values to determine how strongly they espouse the corresponding ideology. Examples of these applications are presented in Section~\ref{sec:eval}.



\begin{table*}[htb]
\caption{Evaluation metrics of clustering result on two Twitter datasets and the Voteview dataset. $^*$Stance Detection is Twitter-specific and not applicable for Voteview. $^\dagger$sPP is developed for signed political bipartite graph and only applicable for Voteview. $^\ddagger$TIMME-Sup is trained in a supervised manner. Stance Detection and TIMME models only support the prediction of users.
}\label{table::clustering}
\centering
\begin{tabular}{llllllll}
\hline
\multicolumn{8}{c}{Dataset: US Election 2020} \\ \hline
  Model Name & User Prec. & User Recall & User F1 & Claim Prec. & Claim Recall & Claim F1 & Purity\\
  \hline
  NMF\cite{al2017unveiling} & 0.4275 & 0.8235 & 0.5628 & 0.4130 & 0.6786 & 0.5135 & 0.6313 \\
  BSMF\cite{yang2020disentangling} & 0.6970 & 0.6866 & 0.6917 & 0.3818 & 0.7778 & 0.5122 & 0.6959 \\
  GCN\cite{kipf2016semi} & 0.5699 & 0.7910 & 0.6625 & 0.3455 & 0.7037 & 0.4634 & 0.6512 \\
  DeepWalk\cite{perozzi2014deepwalk} & 0.9310 & 0.8060 & 0.8640 & \textbf{0.8824} & 0.5556 & 0.6818 & 0.8571 \\
  Stance\cite{darwish2020unsupervised} & \textbf{0.9429} & 0.6226 & 0.7500 & - & - & - & 0.8240 \\
  TIMME-Unsup\cite{xiao2020timme} & 0.9322 & 0.8209 & 0.8730 & - & - & - & 0.7822 \\
  \textbf{InfoVGAE (Ours)} & 0.9333 & \textbf{0.8358} & \textbf{0.8819} & 0.6667 & \textbf{0.8148} & \textbf{0.7333} & \textbf{0.8599} \\
  \hline
  TIMME-Sup\cite{xiao2020timme}$^\ddagger$ & 1.0000 & 0.8333 & 0.9091 & - & - & - & -\\
  \hline
\multicolumn{8}{c}{Dataset: Eurovision 2016} \\ \hline
Model Name & User Prec. & User Recall & User F1 & Claim Prec. & Claim Recall & Claim F1 & Purity\\ \hline
    NMF & 0.3202 & 0.9286 & 0.4762 & 0.3142 & 0.5352 & 0.3960 & 0.7123 \\
    BSMF & 0.5337 & 0.6786 & 0.5975 & 0.2866 & 0.5352 & 0.3733 & 0.7248 \\
    GCN & 0.3113 & \textbf{0.9429} & 0.4681 & 0.2918 & 0.8594 & 0.4356 & 0.7135 \\
    DeepWalk & 0.3028 & \textbf{0.9429} & 0.4583 & 0.2895 & \textbf{0.8867} & 0.4365 & 0.7217 \\
    Stance & 0.4280 & 0.9134 & 0.5829 & - & - & - & 0.6947 \\
    TIMME-Unsup & 0.9513 & 0.7778 & 0.8556 & - & - & - & 0.7865\\
    \textbf{InfoVGAE (Ours)} & \textbf{0.9649} & 0.7857 & \textbf{0.8661} & \textbf{0.8447} & 0.5312 & \textbf{0.6523} & \textbf{0.8842} \\
    \hline
    TIMME-Sup$^\ddagger$ & 0.9907 & 0.7852 & 0.8760 & - & - & - & -\\
    \hline
\multicolumn{8}{c}{Dataset: Voteview$^*$} \\ \hline
Model Name & User Prec. & User Recall & User F1 & Claim Prec. & Claim Recall & Claim F1 & Purity\\ \hline
    NMF & \textbf{0.9952} & 0.9763 & 0.9856 & 0.4957 & \textbf{0.9971} & 0.6622 & 0.8451 \\
    BSMF & 0.9718 & 0.9764 & 0.9741 & 0.4826 & \textbf{0.9971} & 0.6504 & 0.8383 \\
    GCN & 0.4742 & 0.9528 & 0.6332 & 0.4203 & 0.8563 & 0.5639 & 0.6149 \\
    DeepWalk & \textbf{0.9952} & 0.9763 & 0.9856 & 0.4922 & \textbf{0.9971} & 0.6591 & 0.8451 \\
    sPP\cite{akoglu2014quantifying}$^\dagger$ & 0.9718 & 0.9764 & 0.9741 & 0.8427 & 0.8621 & 0.8523 & 0.9430\\
    TIMME-Unsup & 0.9765 & 0.9432 & 0.9595 & - & - & - & 0.8827\\
    \textbf{InfoVGAE (Ours)} & \textbf{0.9952} & \textbf{0.9811} & \textbf{0.9881} & \textbf{0.9878} & 0.9339 & \textbf{0.9601} & \textbf{0.9828} \\
    \hline
    TIMME-Sup$^\ddagger$ & 0.9850 & 0.9924 & 0.9887 & - & - & - & -\\
    \hline
\end{tabular}
\end{table*}



\section{Experiments}
\label{sec:eval}
In this section, we evaluate the performance of the proposed InfoVGAE on mainly three real-world datasets collected from Twitter and the Voteview~\cite{lewis2019voteview} U.S. Congress voting database. We compare the proposed InfoVGAE with $8$ baselines. We demonstrate the versatility of the computed latent belief representation at a number of downstream tasks and compared to prior solutions from the state of the art. Finally, ablation studies are conducted to further demonstrate the effectiveness of design decisions in InfoVGAE. In all the following experiments, the latent space dimension is set as $3$. The experiments are conducted based on Python 3.6.2 and Pytorch 1.7.0 framework, on a device with 12-core CPU and 64GB RAM.

\subsection{Datasets}\label{app:dataset}
Earlier in the paper, we made a distinction between stance detection and polarity classification. We aim to show in the evaluation that our belief embedding algorithm can, in fact, tackle {\em both problems\/} (and improve on the existing state of the art). Thus, we first use two Twitter datasets that are closer to stance detection problems (each data set featuring one broad topic with conflicting views), then include a Voteview dataset with voting records that exemplifies belief embedding (featuring individuals who vote on a large variety of different issues not clearly related to a common topic, but that nevertheless expose societal polarization).

\noindent
\textbf{US Election 2020}
We collected a real-world dataset via the Twitter search API using keywords \textit{\{president, election, trump, biden\}}. A total of $1,149,438$ tweets were collected about the US presidential election from \textit{Dec 4, 2020} to \textit{Dec 23, 2020}. The dataset captures debate about the legitimacy of the election process and includes many opinion statements about Donald Trump, Joe Biden, and related events in their campaign.
Individual tweet cascades (i.e., a tweet and its retweets) were called claims, one per cascade.
We asked human graders to manually label $844$ most retweeted claims for evaluation as either pro-Trump or pro-Biden. Among our labeled claims, there were $237$ claims supporting Trump and $607$ claims supporting Biden.

\noindent
\textbf{Eurovision}
This public dataset is about the annual Eurovision Song Contest~\cite{al2017unveiling}. It was used for polarity detection. The background is that Susana Jamaladinova (Jamala) from Ukraine won the Eurovision 2016 contest with a song named \textit{1944}. This song ignited controversy due to political connotations and possible violations of Eurovision rules. Some users opposed the song quoting Eurovision rules that prevent politically motivated entries. Others applauded Jamala for her rendition of the plight of an ethnic minority, who suffered (presumably by Russian hands) as described in the song. In this dataset, $600$ claims were labeled pro-Jamala and $239$ were labeled anti-Jamala.

\noindent
\textbf{Voteview}
We collected the voting data of the $105^{th}$ Congress (that held office towards the end of the 90s) from the Voteview~\cite{lewis2019voteview} database that documents U.S. Congress voting records. Our collected data contains information on $444$ congressmen from different parties, $1166$ bills, and the voting records. Since most congressmen are Republican or Democrat, we only consider them for polarization analysis. For ground truth, we label the congressmen with their party affiliations, and label the bills with the majority party affiliation of congressmen who voted \textit{Yea}.




\subsection{Baselines}\label{app:baseline}

\noindent\textbf{Non-Negative Matrix Factorization (NMF)}~\cite{al2017unveiling}:~~~~This is an unsupervised approach that uncovers polarization based on factorizing the matrix of users and their claims. Unlike the VGAE, matrix factorization breaks down an observations matrix into an encoder (matrix) and a decoder (matrix) that are {\em both linear\/}.

\noindent\textbf{Belief Structured Matrix Factorization (BSMF)}~\cite{yang2020disentangling}:~~~~This is an enhancement to Non-Negative Matrix Factorization handling situations where different community belief systems partially overlap.

\noindent\textbf{Graph Convolutional Networks (GCN)}~\cite{kipf2016semi}:~~~~The regular GCN is a semi-supervised model that encodes node features and graph structure into representations with a small set of labels. For a fair comparison, we adopt an unsupervised GCN with a softmax classifier after GCN layers for link prediction during training.

\noindent\textbf{Stance Detection (Stance)}~\cite{darwish2020unsupervised}:~~~~An unsupervised stance detection model that uses texts, hashtags, and mentions to build features, and maps users into a low dimensional space with UMAP.


\noindent\textbf{DeepWalk}~\cite{perozzi2014deepwalk}:~~~~This method learns a latent social representation by modeling a series of short random walks. It maps the nodes into a relatively small number of dimensions, capturing neighborhood similarity and community membership.

\noindent\textbf{Signed Polarity Propagation (sPP)}~\cite{akoglu2014quantifying}:~~~~This method represents opinions of individuals with signed bipartite networks and formulates polarity analysis as a classification task. A linear algorithm is proposed to learn the polarity labels exploiting network effects.

\noindent\textbf{TIMME (TIMME-Sup)}~\cite{xiao2020timme}:~~~~TIMME is a supervised multi-task and multi-relational embedding model. TIMME first models the social networks as a heterogeneous graph and jointly trains several link prediction tasks and an entity classification task to extract a latent representation of users. We use TIMME-Sup to refer to the TIMME-hierarchical architecture in the original implementation.

\noindent\textbf{Unsupervised TIMME (TIMME-Unsup)}~\cite{xiao2020timme}:~~~~An unsupervised variant of TIMME-hierarchical without the supervision of entity classification task.

\subsection{Belief Embedding Visualization}
We first visualize the learned representations of InfoVGAE and the baselines in Figure~\ref{fig::embedding}. The users (Twitter accounts and congressmen) and claims (tweets and bills) of view 1 (in blue) are associated with Dim2 axis while those of view 2 (in orange) are associated with Dim3 axis. In the figure, the projection of one point in Dim3 represents its degree of supporting view 1 while the projection in Dim2 represents the degree of supporting view 2. The higher value in each axis means a higher degree of supporting one view. In addition, the projection in Dim1 represents the degree of supporting the other views, which are not often related to the main topics.

In Figure~\ref{fig::embeddingOUR}, we can observe that the representations learned by our method can be easily separated by the diagonal between Dim2 and Dim3. For the baseline methods, however, it is difficult to directly separate the learned representations. In other words, InfoVGAE facilitates polarity classification compared to the baselines.

The Stance Detection method in Figure~\ref{fig::embeddingStance} performs well on the US election dataset, but it does not work well on the Eurovision dataset. We can see from the middle subfigures that some state-of-the-art models such as Stance Detection in Figure~\ref{fig::embeddingStance} and TIMME-Unsup in Figure~\ref{fig::TIMME} produce representations that intersect with each other, which makes it hard to separate them. In addition, the other methods, such as the GCN model in Figure~\ref{fig::embeddingGCN}, DeepWalk in Figure\ref{fig::embeddingDeepWalk}, NMF in Figure~\ref{fig::embeddingNMF} and BSMF in Figure~\ref{fig::embeddingBSMF} do not do well at separating different views in a latent space.

\subsection{Polarity and Stance Detection}
After learning the latent representations, the simplest downstream application is stance and/or polarity detection. While with InfoVGAE classification can be done based on coordinate values, for the compared baselines we apply a K-Means clustering algorithm to predict the polarity of users and claims. The number of clusters is set to $2$, since in all the datasets, the ground-truth label is binary. In this experiment, we conduct clustering separately for users and claims for a fair comparison since some baselines only produce user representation. We use common information retrieval metrics including precision, recall, and F1-score to evaluate the performance of our method. We also calculate the purity metric introduced in~\cite{darwish2020unsupervised,mogotsi2010christopher}. To compute purity, each cluster is assigned to the class which is the most frequent in the cluster. The purity is measured by counting the number of correctly assigned data points and divided by the total number of points $N$. Assume $k$ is the number of clusters, $\mathbf{\Omega}=\{\omega_1, \omega_2, \cdots, \omega_k\}$ is the set of predicted clusters and $\mathbf{C}=\{c_1, c_2, \cdots, c_k\}$ is the set of classes, the purity score is defined as $purity = \frac{1}{N}\sum_{i=1}^{k}\max_j|\omega_i \cap c_j|$, where $c_j$ is the classification which has the maximum count for cluster $\omega_i$. A higher purity represents a cleaner separation of polarized groups.

The evaluation results are shown in Table~\ref{table::clustering}. The results for the Twitter datasets illustrate stance detection (pro-Jamala versus against, pro-Trump versus against). The results for Voteview dataset illustrate polarity separation (Democrats versus Republicans). We observe that InfoVGAE achieves the highest F1-Score and purity on all the three datasets. The reason is that our model can jointly map users and claims into a disentangled latent space that mutually enhances each other. 

DeepWalk has a higher F1 score and purity than most other baselines on the US Election dataset but does not work well on Eurovision, because the Eurovision dataset contains more noise, causing the Deepwalk to generate many redundant clusters. Stance and TIMME-Unsup are state-of-the-art stance detection and social graph representation learning methods. Unlike InfoVGAE, they can only produce the representations of users due to their specific model frameworks. Therefore, we only compare the metrics of users. The Stance Detection model produces a relatively low F1 score. As shown in Figure~\ref{fig::embeddingStance}, the reason is the distance between the two labels is very small, making it hard for the clustering algorithm to separate them. The F1 score of TIMME-Unsup is comparative but slightly lower than the InfoVGAE model. This is also attributed to the drawback of its inexplainable latent space distribution. GCN in Figure~\ref{fig::embeddingGCN} shows a boundary between the points of two different views. However, their spatial distribution is almost the same. This makes cluster separation harder and results in a low F1 score.

Some baselines produce a higher recall or precision but a smaller F1 score, compared with InfoVGAE. The reason is the clustering algorithm is confused by the embedding and mistakenly cluster most points together as one category. In the Voteview dataset, the adjacency matrix represents the voting record of congressmen. It is more dense and contains less noise. Therefore, most models produce a high User F1-score over $0.97$. However, the baselines still cannot achieve a comparable Bill F1-score as InfoVGAE, because their bill representations are less informative and harder to separate.

\begin{table}[htb]
\vspace{-5pt}
\caption{Evaluation results on the PureP dataset~\cite{xiao2020timme}. InfoVGAE only utilizes tweet records. InfoVGAE+Follow further leverages the follow relations. The other models utilize all the additional inputs (node features, follows, likes, etc.).}\label{table::timmedataset}
\centering
\begin{tabular}{lcc}
\hline
  Unsupervised Models & Accuracy & F1 score \\
  \hline
TIMME-Unsup & 0.9860 & 0.9858\\
InfoVGAE (Ours) & 0.9837 & 0.9838\\
\textbf{InfoVGAE+Follow (Ours)} & \textbf{0.9930} & \textbf{0.9931} \\
\hline
Supervised Benchmarks & Accuracy & F1 score \\
\hline
HAN~\cite{wang2019heterogeneous} & 0.9825 & 0.9824\\
TIMME & 0.9825 & 0.9824\\
TIMME-single & 1.000 & 1.000\\
TIMME-hierarchical & 1.000 & 1.000\\
\hline
\end{tabular}
\end{table}

\begin{table}[htb]
\caption{Tweets with the Top-5 highest polarities in US Election dataset, ranked by the coordinate value of corresponding axis, in the latent space of learned belief representation.}
\label{table::caseElection}
\vspace{-5pt}
\footnotesize
\begin{tabular}{|p{0.21\textwidth}|p{0.21\textwidth}|}
\hline
\\[-1.1em]
\multicolumn{1}{|c|}{\textbf{Pro-Trump}} &  \multicolumn{1}{c|}{\textbf{Anti-Trump}} \\ 
\\[-1.1em]
\hline
Democrats would never put up with a Presidential Election stolen by the Republicans! & In all the years I worked in GOP politics, it never occurred to me that there were people sitting in rooms with me that would one day be in favor of overturning a Presidential election, making the loser America's first dictator. \\ \hline
The only thing more RIGGED than the 2020 Presidential Election is the FAKE NEWS SUPPRESSED MEDIA. & Pence, any comment yet on who won the 2020 Presidential election, or do you need still more time to think?  \\ \hline
@senatemajldr and Republican congressmen have to get tougher, or you won't have a Republican Party anymore. & Losing the 2020 Presidential election, Trump threatened to "take it to the Supreme Court," but tonight he found out that the Supreme Court is not some kind of walk-up bank teller.           \\ \hline
The answer to the Democrat voter fraud is not to stay at home - that's what Pelosi and Schumer want you to do. & Trump's insane tweets now carry a very different label: "Election officials have certified Joe Biden as the winner of the U.S. Presidential election." \\ \hline
TX presidential electors cast all 38 votes for Trump \& Pence... they also voted to "condemn" SCOTUS tossing out TX election lawsuit. & Trump just posted this map, allegedly depicting the results of the 2020 presidential election.  It's a fraudulent map. \\
\hline
\end{tabular}
\vspace{-5pt}
\end{table}

We also compare the result of InfoVGAE with a supervised model. InfoVGAE produces a very comparative result with TIMME-Sup, while InfoVGAE is an unsupervised method. In the Voteview dataset, the gap of F1 score is only $0.06\%$. In US Election and Eurovision datasets, the gap is also narrowed into  $1.86\%$. It's reasonable that supervised methods outperform unsupervised ones. However, the evaluation result of InfoVGAE is closely approaching the upper bound of all unsupervised methods, with a special design and control for latent space distribution.

In addition, we evaluate the result of InfoVGAE in the PureP dataset published in~\cite{xiao2020timme}, where many additional inputs (\textit{node features, follows, likes, mentions, and replies}) are available. The accuracy and F1 score of user classification are shown in Table~\ref{table::timmedataset}. The proposed InfoVGAE is designed to work in a \textit{“minimalist”} scenario where additional information is not available. It only requires tweet records as the inputs. However, it's also compatible with additional inputs. To integrate the follow information in PureP dataset, we add edges between the following-follower user pairs in the input heterogeneous graph. By integrating only the users' follow information, the InfoVGAE+Follow methods achieve the highest accuracy of $0.9930$ and the highest F1 score of $0.9931$ among unsupervised baselines. It also achieves very close and even better performance compared with supervised models.

\subsection{Case Study of Stance Separation}
Many existing stance detection models only support the stance evaluation of users, such as Stance Detection~\cite{darwish2020unsupervised} and TIMME~\cite{xiao2020timme}, whereas the design of InfoVGAE enables us to further separate the stances of claims made by users. We show the top 5 claims separated by stance by our unsupervised algorithm (to get a feel for the data at hand). As mentioned above, each axis in the disentangled latent space produced by InfoVGAE is associated with a different ideology. Based on this observation, we sort and rank the claims by their coordinates on each axis and report (for illustration) the top-5 claims (with the largest coordinate) on each axis. Results are shown in Table~\ref{table::caseElection} (for the US election) and Table~\ref{table::caseEurovision} (for Jamala). Column labels are added manually.

\begin{table}[htb]
\caption{Tweets with the Top-5 highest polarities in Eurovision dataset, ranked by the coordinate value of corresponding axis, in the latent space of learned belief representation.}
\label{table::caseEurovision}
\vspace{-5pt}
\footnotesize
\begin{tabular}{|p{0.21\textwidth}|p{0.21\textwidth}|}
\hline
\\[-1.1em]
\multicolumn{1}{|c|}{\textbf{Pro-Jamala}} &  \multicolumn{1}{c|}{\textbf{Anti-Jamala}} \\ 
\\[-1.1em]
\hline
RT @jamala: Thank you for your love! \#jamala \#eurovision \#jamala1944 \#eurovision\_ukraine  \#cometogether & Jamala's 1944: Song for Nazi.         \\ \hline
Incredible performance by \#Jamala, giving Crimean Tatars, suffering persecution \& abuse, reason to celebrate.  & I must say I feel a little sorry for @jamala, from the start simply a tool in the West's \#CrimeanTatars campaign.          \\ \hline
President awarded @jamala title of the People's Artist of Ukraine. & if Jamala's singing of "Our Crimea", a totally non-political song, wasn't against the rules - why is every video of it removed now?         \\ \hline
\#CantStopTheFeeling\#Eurovision. Congrats @jamala \#Ukraine!! & Ukraine's Eurovision winner, Jamala, is so angry with Russia that she appeared at Sochi's New Year party (\$\$)           \\ \hline
This scene will give me goosebumps until the day I die. Thanks for such a masterpiece @jamala. & @jtimberlake @Eurovision @jamala The day Eurovision REALLY went political. What a shame \#Eurovision.  \\ \hline
\end{tabular}
\vspace{-5pt}
\end{table}

Now how, in Table~\ref{table::caseElection}, all \textit{Pro-Trump} tweets are indeed supporting Trump and the Republican Party, and advocating that people vote for Trump. All \textit{Anti-Trump} tweets are condemning and/or post sarcasm on Trump. In Table~\ref{table::caseEurovision},  \textit{Pro-Jamala} tweets are expressing gratitude and congratulate Jamala. The \textit{Anti-Jamala} tweets criticize Jamala's song as political and a violation of the rules of the Eurovision contest.

We notice that sarcasm and irony are common in both datasets, which are difficult to be understood by a machine (and even humans without context). For example, if we read the first \textit{Pro-Trump} tweet semantically, it seems to accuse Republicans of stealing an election. This tweet is actually posted by Trump himself to emphasize Republicans would not allow an election to be (allegedly) stolen either. In the second \textit{Anti-Trump} tweet, the person asks a sarcastic question, with the implication that the answer should obvious. In the fourth \textit{Anti-Jamala} tweet, the user posts irony with "\$", suggesting that Jamala appeared at the new year party in Sochi, Russia for money, while she is a Ukrainian. These examples illustrate the benefits of InfoVGAE's language-agnostic embedding.

\subsection{Stance Prediction}
\begin{figure}[!ht]
	\centering
	\includegraphics[width = 0.99\linewidth]{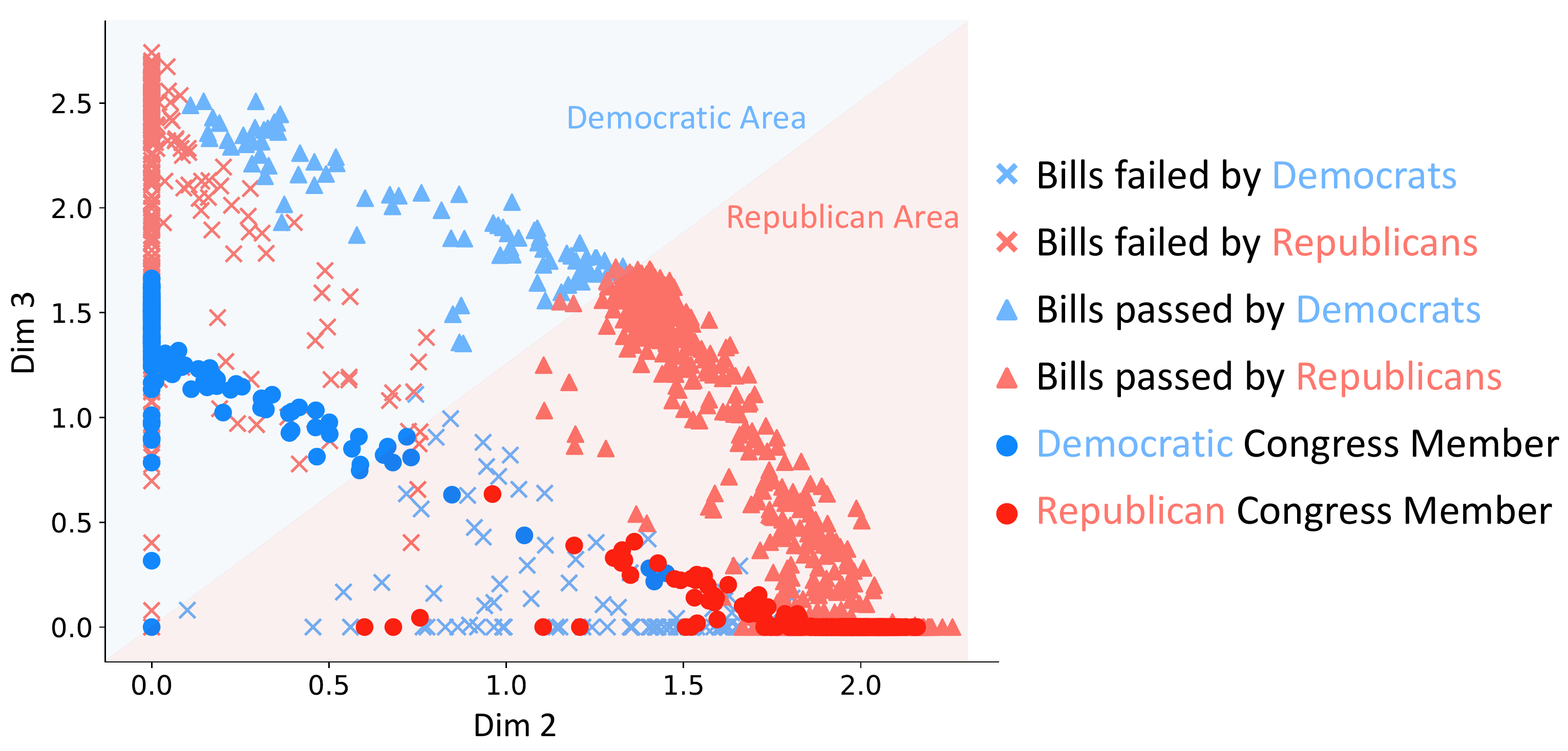}
	\caption{Learned Representations in Voteview dataset.
	We projected the 3D representations into 2D in order to easily understand the results of polarity detection by InfoVGAE. The colors and shapes of data points represent ground-truth.}
	\label{fig::bill}
	\vspace{-10pt}
\end{figure}

We do not actually present detailed results on stance {\em prediction\/} here, but rather present evidence that it should be possible to predict stance from the embedding. Figure~\ref{fig::bill} is a 2D projection of the InfoVGAE's latent representations of congressmen and bills. It shows ground truth on passed and failed bills as well as ground truth on the party responsible for passing or failing them. It also shows the ground truth party affiliation of Congress members. The diagonal separates the two belief systems. Note how most bills above the diagonal (in the Democrat space) are either passed by Democrats or failed by Republicans. Similarly, most bills below the diagonal (in the Republican area) are either passed by Republicans or failed by Democrats. The figure shows that the latent representation of bills learned by InfoVGAE indeed predicts the parties which will vote for/against them.

\subsection{Ideology Analysis}
The Voteview database provides an ideology value~\cite{boche2018new} generated by DW-NOMINATE~\cite{mccarty1997income} algorithm, which represents the static ideological position of each Congress member across the course of their career. This ideology value is calculated based on large amounts of data in history and can be used as the benchmark for congressman's ideological leanings. In this scoring system, the ideology value is positive for Republican congressmen and negative for Democratic congressmen. A larger absolute value means a more deeply entrenched position.

\begin{table}[htb]
\vspace{-5pt}
\caption{Ideology evaluation in Voteview dataset. We report the overall Kendall correlation as well as the Kendall within the Republican (R.) and Democratic (D.) parties. A higher Kendall correlation and cosine similarity represents a better match between ideology and the predicted polarity.}\label{table::ranking}
\centering
\vspace{-5pt}
\begin{tabular}{lcc}
\hline
  Model Name & Overall~/~R.~/~D.~Kendall & Cosine Similarity \\
  \hline
  NMF & 0.7168~/~0.6579~/~0.6637 & 0.6739\\
  BSMF & 0.7751~/~0.7186~/~0.7059 & 0.9650\\
  GCN & 0.4934~/~0.4571~/~0.4586 & 0.0263\\
  DeepWalk & 0.7761~/~0.7039~/~0.7171 & 0.9655\\
  TIMME-Unsup & 0.7554~/~0.7022~/~0.6983 & 0.9478\\
  \textbf{InfoVGAE} & \textbf{0.7876}~/~\textbf{0.7207}~/~\textbf{0.7287} & \textbf{0.9688} \\
 \hline
\end{tabular}
\vspace{-5pt}
\end{table}

\begin{figure}[!ht]
	\centering
	\includegraphics[width = 0.9\linewidth]{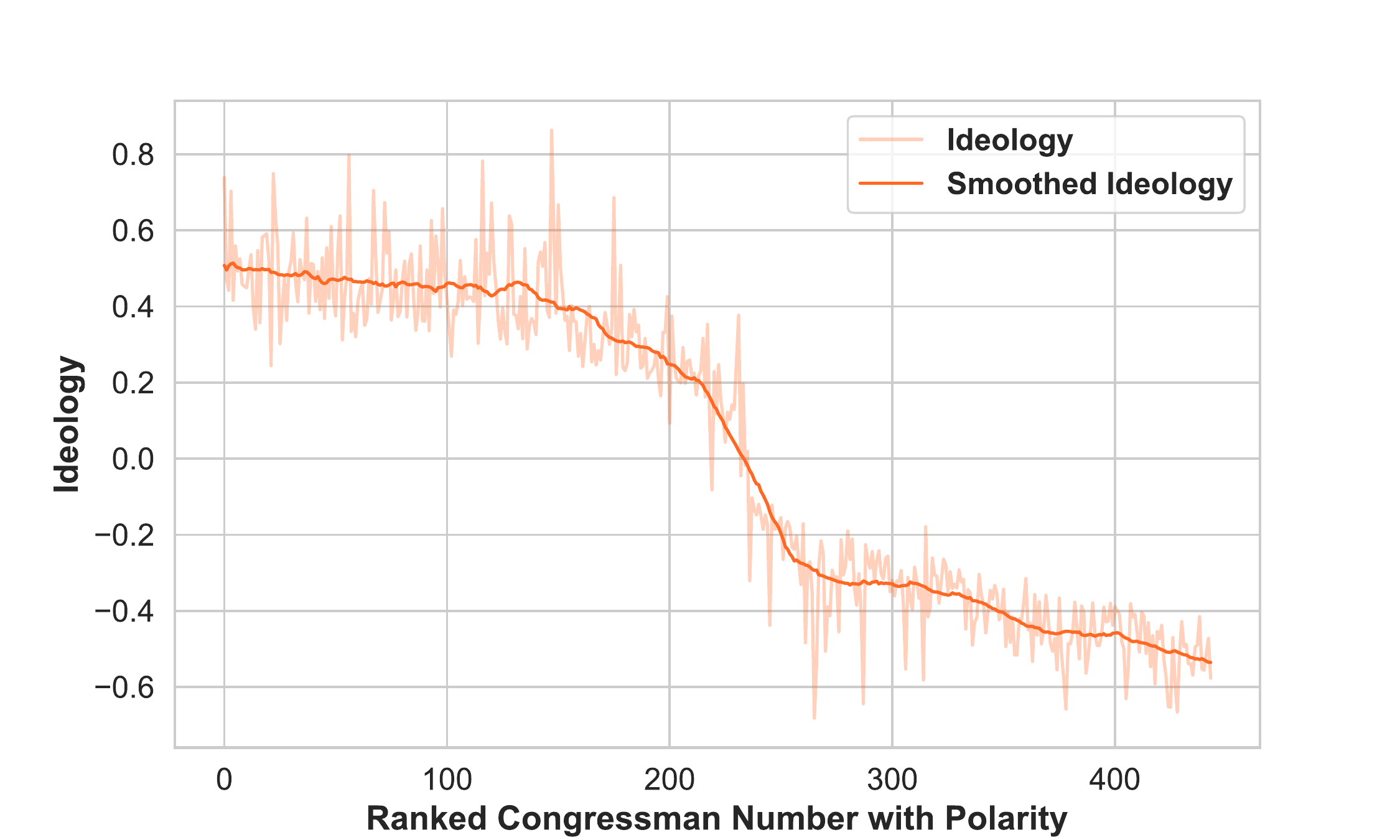}
	\caption{Ideology of ranked Congressmen. The Smoothed ideology is smoothed with a slide window. The overall trending of the ideology is monotonic, which means the ranking predicted by InfoVGAE is consistent with the ideology.}
	\vspace{-5pt}
	\label{fig::ideology}
\end{figure}

Ideally, the produced polarity value by InfoVGAE (i.e., the projection of a point on its dominant ideology axis) should be strongly correlated with the ground-truth ideology of congressmen. To evaluate the correctness of the polarity ranking produced by our InfoVGAE model, we first visualize the ideology of ranked congressmen in Figure~\ref{fig::ideology}. The horizontal axis is the order of congressmen ranked with their polarity value. In the figure, clear correlations are seen with the ground-truth ideology value. Quantitatively, we use Kendall Rank Correlation Coefficient within and across parties to evaluate the correlation between the ranking sequences of polarity and ideology. We also use Cosine Similarity to evaluate the similarity between polarity and ideology values. We calculate the polarity of baselines with the clustering-based method introduced in Section~\ref{sec::unsupervised}. The result is shown in Table~\ref{table::ranking}. InfoVGAE achieved the highest Kendall correlation and cosine similarity, without even running clustering algorithms.

Next, we show the top $5$ congressmen of the highest latent value of polarity (according to Figure~\ref{fig::bill}) in Table~\ref{table::congressmen}. Those individuals espouse the most extreme positions. We look up their ground truth ideology score, showing that they are indeed outliers compared to the average ideology score for the party, shown in the last row. This table offers further intuition into the quality of ranking by latent polarity computed by InfoVGAE.

\begin{table}[htb]
\caption{Top-5 Congressmen with the highest polarity. Ideology represents ground-truth of static ideological position.}
\vspace{-5pt}
\small
\label{table::congressmen}
\begin{tabular}{|c|c|c|c|}
\hline
Democratic & Ideology & Republican & Ideology\\
\hline
\hline
 OLVER, John Walter& $-0.577$ & CRANE, Philip Miller & $0.739$ \\
 \hline
 VENTO, Bruce Frank & $-0.472$ & PAXON, L. William & $0.472$ \\
 \hline
 MINK, Patsy T. & $-0.513$ & BRYANT, Ed & $0.442$ \\
 \hline
 WOOLSEY, Lynn C. & $-0.556$ & STUMP, Robert Lee & $0.703$\\
 \hline
 MILLER, George & $-0.552$ & HASTINGS, Doc & $0.416$\\
 \hline
 \textit{Democratic Average} & $-0.376$ & \textit{Republican Average} & $0.402$ \\
 \hline 
\end{tabular}
\vspace{-5pt}
\end{table}


\subsection{Ablation Studies}\label{sec::ablation}
We further conduct ablation studies to explore the impact of proposed modules on polarity detection. We keep all other experimental settings unchanged except for the ablation module. The experimental results are shown in Table~\ref{table::ablation}.

\noindent\textbf{Effect of Total Correlation Module:}~~~~
We remove the discriminator for total correlation regularization and remove the total correlation term in our objective function. In this way, the independence of axes in the learned embedding space is no longer guaranteed. This limits the ability of InfoVGAE to compute an informative representation and therefore reduces the F1 scores.

\noindent\textbf{Effect of Rectified Gaussian Distribution:}~~~~
We apply a general Gaussian distribution instead of the rectified Gaussian distribution to learn the distribution of latent representations. Therefore, the values of latent variables become any real numbers rather than non-negative ones. We can observe from Table~\ref{table::ablation} that the performance of InfoVGAE with the general Gaussian distribution is reduced.


\noindent\textbf{Effect of KL Divergence Control:}~~~~
Next, we study the impact of the PI control algorithm on the performance of polarity detection. We remove the PI control algorithm in the InfoVGAE. As illustrated in Table~\ref{table::ablation}, its F1 scores for users and tweets decrease on all the datasets, especially for the Voteview dataset. This is attributed to the dense graph of Voteview dataset, since the voters usually vote for most of the bills. The dense graph leads to an unstable KL divergence during the training process. The KL control module helps control the KL divergence within a reasonable range, therefore the learned representations become more informative.



\noindent\textbf{Effect of Joint Learning of Tweets and Users:}~~~~
InfoVGAE constructs BHIN containing both user and claim nodes to jointly learn their embedding. We conduct an ablation study to test its effectiveness by separately building two graphs containing users or claims, and learning embeddings respectively. The evaluation metrics of the separate-learning version are $2\%$-$6\%$ lower than joint learning.

Based on the above ablation studies, we conclude that the total correlation module, non-negative latent space, PI control algorithm, and joint learning with BHIN plays an important role to learn a meaningful and disentangled embedding for polarity detection.

\begin{table}[!ht]
\caption{Evaluation results of ablation studies. The first four rows of each dataset show the results after removing the total correlation module, the PI controller, replaceing the rectified Gaussian, or separately learning the embeddings}\label{table::ablation}
\centering
\vspace{-3pt}
\begin{tabular}{llll}
\hline
\multicolumn{4}{c}{Dataset: US Election 2020} \\ \hline
  Model Name & User F1 & Tweet F1 & Purity \\
  \hline
  Without Total Correlation & $0.8780$ & $0.7273$ & $\textbf{0.8660}$\\
  Without Rectified Gaussian & $0.8594$ & $0.4938$ & $0.7753$ \\
  Without KL Control & $0.8358$ & $0.6545$ & $0.8097$\\
  Separate-Learning & $0.8615$ & $0.7111$ & $0.8597$\\
  \textbf{InfoVGAE} & $\textbf{0.8819}$ & $\textbf{0.7333}$ & $0.8599$ \\
  \hline
\multicolumn{4}{c}{Dataset: Eurovision 2016} \\ \hline
Model Name & User F1 & Tweet F1 & Purity\\
    \hline
    Without Total Correlation & $0.8625$ & $0.5100$ & $0.8196$ \\
    Without Rectified Gaussian & $0.6209$ & $0.5093$ & $0.7144$ \\
    Without KL Control & $0.8506$ & $0.5045$ & $0.8175$ \\
    Separate-Learning & $0.8497$ & $0.6449$ & $0.8249$\\
    \textbf{InfoVGAE} & $\textbf{0.8661}$ & $\textbf{0.6523}$ & $\textbf{0.8842}$ \\
    \hline
\multicolumn{4}{c}{Dataset: Voteview} \\ \hline
Model Name & User F1 & Tweet F1 & Purity \\
    \hline
    Without Total Correlation & $\textbf{0.9881}$ & $0.9518$ & $0.9807$ \\
    Without Rectified Gaussian & $\textbf{0.9881}$ & $0.9309$ & $0.9751$ \\
    Without KL Control & $0.9858$ & $0.6578$ & $0.8440$\\
    Separate-Learning & $0.9791$ & $0.9218$ & $0.9659$\\
    \textbf{InfoVGAE} & $\textbf{0.9881}$ & $\textbf{0.9601}$ & $\textbf{0.9828}$ \\
    \hline
\end{tabular}
\vspace{-5pt}
\end{table}

\section{Related Work}\label{sec:relatedwork}

The past few years witnessed a large surge of work on stance detection and polarity classification~\cite{stefanov2020predicting,mohtarami2019contrastive,badawy2019falls,darwish2017improved,hasan2014you,bermingham2011using,magdy2016isisisnotislam,jiang2021social}. 

While some stance detection relied on sentiment analysis~\cite{bermingham2011using}, most studies framed the stance detection problem as a supervised classification problem~\cite{hasan2014you,darwish2017improved,jiang2021social,magdy2016isisisnotislam,li2019senti2pop,kuccuk2020stance} or a transfer learning problem~\cite{mohtarami2019contrastive} that correlates text features and user network neighborhoods with stance~\cite{darwish2017improved,stefanov2020predicting}. Multi-target stance prediction explored correlations between different stances (e.g., with respect to election candidates)~\cite{sobhani2017dataset}. Traditional machine learning algorithms~\cite{volkova2014inferring,barbera2015follow,barbera2015understanding}, such as SVM, Random Forest, and Bayesian estimators, were used. For example, Da Silva et al.~\cite{da2014tweet} developed an ensemble method that combines RF, SVM, and LR to improve the classification accuracy.

With advances in deep neural networks, recent work applied deep learning models to detect polarity by mapping people's systems of belief into a latent space~\cite{xiao2020timme,tang2017multiple,jiang2021social,siddiqua2019tweet,rashed2020embeddings,umer2020fake}. For example, Jiang et al.~\cite{jiang2021social} developed a weakly supervised model, Retweet-BERT, to predict the polarity of users on Twitter based on network structures and content features. Xiao et al.~\cite{xiao2020timme} proposed a multitask multi-relational embedding model to predict users' ideology using graph convolutional networks (GCN). However, the problem of supervised learning approaches is that they require human-annotated data, which is costly and time-consuming.

To deal with this issue, some work adopted unsupervised learning models for stance and/or polarity detection~\cite{akoglu2014quantifying,garimella2018political,trabelsi2018unsupervised,stefanov2020predicting,jang2018explaining}. Unsupervised solutions were developed for clustering users by stance or viewpoint~\cite{trabelsi2018unsupervised}. For example, Jang et al.~\cite{jang2018explaining} proposed a probabilistic ranking model to detect the stance of users in two communities. In addition, researchers~\cite{trabelsi2018unsupervised} developed a purely unsupervised Author Interaction Topic Viewpoint model (AITV) for polarity detection at the post and the discourse levels. However, these methods do not focus on belief embedding.

Generalizing from stance classification problems, some work explored ideology as a variable that changes within a range~\cite{barbera2015follow}. It was postulated that human stances on issues can be predicted from a low-dimensional model~\cite{bogdanov2013social,yang2020disentangling,ange2018semi,darwish2020unsupervised}. Ange et al.~\cite{ange2018semi} developed a semi-supervised deep learning model based on multi-modal data for polarity detection. After that, Darwish et al.~\cite{darwish2020unsupervised} adopted an unsupervised stance detection model that maps users into a low dimensional space based on their similarity. These methods, however, are mostly focused on either user polarity or statement polarity, but usually not both jointly. Extending this view, we develop an unsupervised belief representation learning model that jointly learns the latent representations of users and claims in the same space, thereby improving user and claim polarity identification. Importantly, we think of unsupervised belief representation learning as a {\em separable problem\/} from the downstream application task. Thus, we show (in the evaluation) how the same approach is trivially applied to stance detection, stance prediction, and polarity separation, among other ideology-related analysis tasks in polarized settings.

\section{Conclusion}
\label{sec:conclude}
In this paper, we propose an Information-Theoretic Variational Graph Auto-Encoder (InfoVGAE) for belief representation learning in an unsupervised manner. It constructs a bipartite heterogeneous graph from the interaction data and jointly learns the belief embedding of both users and their claims in the same latent space. InfoVGAE includes three modules to better disentangle the latent space and learned more informative representations for downstream tasks. It adopts the rectified Gaussian distribution to create an orthogonal latent space, which assigns the belief systems into axes. A KL divergence PI controller is applied to balance the trade-off between reconstruction quality and disentanglement. In addition, a total correlation regularizer is proposed to learn a series of statistically independent latent dimensions. Experimental results show that the proposed InfoVGAE outperforms the existing unsupervised polarity detection methods, and achieves a highly comparable F1 score and purity result with supervised methods.

\begin{acks}
This work was conducted in part under DARPA award HR001121C0165, and in part under DoD Basic Research Office award HQ00342110002.
\end{acks}


\bibliographystyle{ACM-Reference-Format}
\bibliography{main}


\end{document}